\documentclass[twocolumn,floats,floatfix,prd,superscriptaddress,nofootinbib]{revtex4-2}
\usepackage[english]{babel}

\usepackage{graphicx,epsfig}
\usepackage{float}

\usepackage{amssymb,amsmath,amsthm,amsfonts}
\usepackage{bm}
\usepackage[inline]{enumitem}
\usepackage{tensor}
\usepackage[linktocpage]{hyperref}
\usepackage[caption=false]{subfig}
\usepackage[usenames,dvipsnames]{xcolor}
\usepackage{url}
\usepackage[normalem]{ulem}
\usepackage{xspace}
\usepackage{comment}
\usepackage{cancel}
\usepackage{multirow}
\usepackage{dcolumn}
\usepackage{diagbox}

\newcommand{\appropto}{\mathrel{\vcenter{
  \offinterlineskip\halign{\hfil$##$\cr
    \propto\cr\noalign{\kern2pt}\sim\cr\noalign{\kern-2pt}}}}}

\makeatletter
\def\l@subsubsection#1#2{}
\def\l@acknowledgements#1#2{}
\makeatother

\newcommand{\sapienza}{Dipartimento di Fisica, Sapienza Università 
	di Roma, Piazzale Aldo Moro 5, 00185, Roma, Italy}
\newcommand{\infn}{INFN, Sezione di Roma, Piazzale Aldo Moro 2, 00185, Roma, Italy}
\newcommand{\aei}{Max Planck Institute for Gravitational Physics (Albert Einstein Institute), Am Mühlenberg 1, 14476 Potsdam, Germany}

\begin{document}
\title{Challenging the Weak Cosmic Censorship with Phantom Fields}

\author{Giovanni Caridi}
\affiliation{\sapienza}
\affiliation{\infn}

\author{Fabrizio Corelli}
\affiliation{\aei}
\affiliation{\sapienza}
\affiliation{\infn}

\author{Paolo Pani}
\affiliation{\sapienza}
\affiliation{\infn}

\begin{abstract}
Penrose's weak cosmic censorship conjecture asserts that spacetime singularities produced by gravitational collapse are generically hidden behind event horizons, thus preventing them from causally influencing distant observers and preserving the predictability of the exterior region. In this work, we probe this conjecture in a setup that deliberately violates one of its central assumptions --~the dominant energy condition~-- by considering the spherical collapse of a \emph{phantom} scalar field with negative energy density. 
In principle, such a field could produce a Schwarzschild geometry with \emph{negative} mass and therefore no event horizon.
Our aim is to assess whether, once the dominant energy condition is abandoned, the fully coupled evolution of matter and geometry can dynamically generate or expose naked singularities, thereby probing the robustness of cosmic censorship. 
To this end, we perform high-accuracy numerical relativity simulations based on fourth-order finite-difference schemes.
Starting from smooth, asymptotically flat initial data representing regular phantom scalar wave packets, we follow their fully nonlinear evolution through collapse or dispersion.
While an ordinary (positive-energy) scalar field exhibits the standard Choptuik critical behavior at the threshold of black-hole formation, the phantom field displays qualitatively different dynamics. For all amplitudes considered, we find no evidence for trapped surfaces, naked singularities, or alternative stationary end states.
Instead, the phantom scalar field disperses in all the configurations studied. Within the parameter space explored, we find no dynamical violation of cosmic censorship.
\end{abstract}

\maketitle

\section{Introduction}

The notion of a \emph{singularity} in General Relativity represents one of the most subtle and conceptually challenging aspects of the theory.
As emphasized by Geroch~\cite{Geroch:1968ut}, 
the familiar intuition of a singular point originates from classical field theories formulated on a fixed background spacetime, where certain physical quantities diverge while the geometric arena remains well defined. 
In classical electrodynamics, for instance, the electric field of a point charge diverges at the source, yet the background spacetime itself remains unaffected. In General Relativity, however, this picture breaks down: the metric is itself a dynamical field, and there is no external geometric structure relative to which singular behavior can be defined. In the absence of a fixed background, the very meaning of singularity becomes more subtle and less intuitive.

For this reason, no universally accepted definition of singularity exists in General Relativity. Historically, two main characterizations have been employed. One identifies singularities with the divergence of curvature invariants, such as the Kretschmann scalar~\cite{Henry:1999rm}. The other, more geometrically robust, definition characterizes a spacetime as singular if it is \emph{geodesically incomplete}~\cite{Misner:1963fr}. In the formulation adopted by Wald~\cite{Wald:1984rg} (see also~\cite{Hawking:1966sx,Hawking:1973uf,Steinbauer:2022hvq,Landsman2021,Penrose:1964wq}), a spacetime is singular if it contains at least one timelike or null geodesic that cannot be extended to arbitrary values of its affine parameter. Physically, this corresponds to a freely falling observer (or photon) whose worldline ends after a finite proper time (or affine parameter), signaling a breakdown of predictability. Even this definition, however, is not entirely free from ambiguities, as it may depend on the differentiable structure or on possible extensions of the spacetime~\cite{Geroch:1968ut}.

A decisive step forward was made by Penrose~\cite{Penrose:1964wq}, who demonstrated that singularities are not artifacts of symmetry but arise generically under gravitational collapse. When a sufficiently massive body can no longer support itself against its own gravity, collapse leads inevitably to geodesic incompleteness. This insight culminated in the singularity theorems of Hawking and Penrose~\cite{Hawking:1973uf}, which show that, under suitable energy and causality conditions, singularities are unavoidable in broad classes of physically relevant spacetimes, including cosmological models and collapsing configurations.

The existence of singularities, however, raises a profound issue. In principle, a singularity formed during collapse need not be hidden behind an event horizon. If visible to distant observers (i.e., if \emph{naked}), such singularities would signal a breakdown of determinism, since classical evolution would cease to be predictable beyond their formation. 
Some solutions of Einstein's equations (e.g.\ certain extremal configurations, dynamical instabilities, and collapse of exotic matter) have been shown to admit naked singularities under specific conditions~\cite{Penrose:1973um,Christodoulou:1994hg,Harada:1999jf,Ori:1987hg,Martin-Garcia:2003xgm,Joshi:1993zg}. 

To address this threat to predictability, Penrose formulated the \emph{weak cosmic censorship conjecture}~\cite{Penrose:1969pc}, which asserts that singularities arising from the gravitational collapse of physically reasonable matter --~obeying the dominant energy condition and evolving from regular initial data~-- are generically hidden within event horizons. Despite decades of effort, this conjecture remains unproven in full generality. A variety of partial results support its validity in restricted settings~\cite{Penrose:1999vj,Christodoulou:1999iwb,Hubeny:1998ga,Semiz:2005gs}. Counterexamples are known in higher dimensions~\cite{Lehner:2010pn,Figueras:2022zkg}, or for test particles~\cite{WALD1974548,Hubeny:1998ga,Jacobson:2010iu} only when neglecting backreaction~\cite{Hubeny:1998ga,Barausse:2010ka,Colleoni:2015afa}.

Scalar-field collapse has played a central role in this debate. Christodoulou's analytical studies demonstrated that spherically symmetric self-gravitating scalar fields can lead either to black-hole formation or, for finely tuned initial data, to the development of naked singularities~\cite{Christodoulou:1999iwb,Christodoulou:2008nj}. Although such naked solutions are unstable under generic perturbations, these results highlight the delicate boundary between collapse outcomes. 

Numerical relativity simulations~\cite{Baumgarte:2010ndz} enabled fully nonlinear investigations of collapse, beginning with Choptuik's discovery of critical phenomena~\cite{Choptuik:1992jv}. Subsequent studies extended these analyses to massive, complex, charged, and self-interacting scalar fields, as well as to perfect fluids~\cite{Brady:1997fj,Choptuik:2004ha,Hod:1996ar,Gundlach:1997wm,Evans:1994pj,Gundlach:1997wm}.

In this work, we deliberately relax one of the key assumptions underlying the weak cosmic censorship conjecture, namely the dominant energy condition. We consider a \emph{phantom} scalar field, described by an inverted-sign kinetic term and therefore characterized by negative energy density. Such fields violate the standard energy conditions~\cite{Hawking:1973uf}  in a controlled manner and provide a useful theoretical laboratory to probe the robustness of horizon formation. 
Indeed, a collapsing phantom field in flat spacetime might in principle form a Schwarzschild metric with \emph{negative} mass. Although such geometry is still asymptotically flat, it is gravitationally repulsive at large radii, unlike the usual positive-mass Schwarzschild spacetime. In the corresponding static continuation, the absence of an event horizon leaves the curvature singularity at $r=0$ naked.

Here we focus on a minimal and clean setup: the spherical collapse of a massless phantom scalar field in asymptotically flat spacetime. Unlike canonical scalar collapse, where attractive gravity competes with dispersion and leads to well-known critical behavior, the inverted kinetic term induces an effectively repulsive gravitational response. This raises a natural question: can such a violation of the dominant energy condition\ prevent horizon formation while still allowing singular structures to develop, thereby challenging the weak cosmic censorship conjecture?
At first glance, one would expect the repulsive character of the phantom field to prevent matter from being compressed into a region small enough to form a singularity. However, due to the nonlinear nature of the gravitational interaction, this expectation is not guaranteed a priori, and a fully nonlinear analysis is required to determine the possible outcomes of the evolution. In particular, sufficiently large and localized negative-energy configurations could in principle undergo a transient strong-field phase leading to singular behavior, or settle into exotic configurations. Phantom fields have indeed been considered in the literature as sources for wormholes and other exotic solutions~\cite{Ellis:1973yv,Martinez:2020hjm}. It is therefore interesting to investigate whether such configurations can arise dynamically from regular initial data during spherical collapse.

To answer this question, we perform high-accuracy numerical relativity\ simulations of smooth, localized wave packets, systematically varying their amplitude and width. Rather than searching for critical scaling laws, our primary goal is to characterize qualitatively the possible dynamical regimes: dispersion, bounce, horizon or other regular end-state formation, or the emergence of naked singularities or other exotic solutions. This setup provides a fully nonlinear and controlled framework to investigate how violations of the energy conditions affect gravitational collapse and to assess the robustness of cosmic censorship beyond its standard assumptions.

Despite being conceptually simple, this setup has received little attention in the literature and, to the best of our knowledge, has only been considered in~\cite{Nakonieczna:2012in,Nakonieczna:2013hs}. In that work, the collapse of scalar fields was studied within an Einstein-Maxwell-dilaton model allowing for phantom fields. Among the configurations analyzed, the case involving only a phantom dilaton coupled to gravity was mentioned to evolve toward a regular spacetime with the scalar field dispersing at infinity. 

In this work we revisit and extend this scenario by systematically exploring the parameter space of the initial wave packet. Our goal is to characterize the dynamics more broadly, investigate the possible formation of naked singularities or other exotic outcomes, and assess the robustness of the behavior previously reported in~\cite{Nakonieczna:2012in,Nakonieczna:2013hs}.

The remainder of this manuscript is organized as follows. In Sec.~\ref{sec:setup} we present the theoretical framework and numerical implementation. In Sec.~\ref{sec:resultscanonical} we discuss the collapse of a canonical scalar field, serving as a benchmark. In Sec.~\ref{sec:resultsphantom} we analyze the phantom case and characterize the resulting spacetime dynamics. Our conclusions are summarized in Sec.~\ref{sec:conclusions}. Appendix~\ref{app:convergence} reports convergence tests of the numerical code and Appendix~\ref{app:KreissOligerImpact} discusses the impact of the dissipation operator on the numerical evolution.

\section{Numerical Setup} \label{sec:setup}
Our code simulates the spherical collapse of wave packets of a scalar field in an initially flat background using Schwarzschild-like coordinates. We start this section by introducing the model and the system of equations we integrate numerically, and then describe the implementation of our code. Here and in the rest of the paper we adopt geometrized units, setting $G = c = 1$. We express all dimensional quantities in units of an arbitrary mass scale $\tilde{M}$.

\subsection{Evolution equations and constraints}
We consider the spherically symmetric Einstein-Klein-Gordon system for a real scalar field $\xi$. The action is
\begin{equation}
\label{eq:action_collapse}
    S=\frac{1}{16\pi}\int_\Omega \sqrt{-g}\Bigr[R+C(\nabla_\mu \xi) (\nabla^\mu \xi)\Bigr]d^4x,
\end{equation}
where \(R\) is the Ricci scalar, \(\xi\) is the scalar field, $g_{\mu\nu}$ is the spacetime metric, and \(C\in\{+1,-1\}\) determines the sign of the kinetic term: \(C=+1\) for a phantom field, and \(C=-1\) for a canonical scalar field. Varying~\eqref{eq:action_collapse} with respect to \(g_{\mu\nu}\) and \(\xi\) we obtain the field equations
\begin{align}
    &G_{\mu\nu}=R_{\mu\nu}-\frac{1}{2}g_{\mu\nu}R=T_{\mu\nu}, \label{eq:fields1}\\  
    &\Box \xi = 0,   \label{eq:fields3}
\end{align}
where $\Box = \nabla_\mu \nabla^\mu$ and the stress-energy tensor contains only the contribution of the scalar field:
\begin{equation}\label{eq:Tmunu_collapse}
T_{\mu\nu}
= C\left(-2(\nabla_\mu\xi)(\nabla_\nu\xi)
+\,g_{\mu\nu}\,\nabla_\rho\xi\,\nabla^\rho\xi\right).
\end{equation}
We express the line element using Schwarzschild-like coordinates:
\begin{equation}\label{eq:metric}
ds^2=-e^{2\alpha(t,r)}\,dt^2+e^{2\beta(t,r)}\,dr^2+r^2\,d\Omega_2^2,
\end{equation}
where $\alpha(t,r)$ and $\beta(t,r)$ are metric functions that depend only on coordinate time and areal radius. While this choice makes the numerical implementation simple, it produces a coordinate singularity at the horizon when a black hole forms. To handle this, we halt the evolution upon horizon detection by introducing diagnostic thresholds on $g_{tt}$ and on quasi-local quantities.

For the scalar field, we introduce the auxiliary variables
\begin{align}
    \Theta &= \partial_r \xi,  \\
    \Pi &= n^\mu \nabla_\mu \xi = e^{-\alpha}\,\partial_t \xi,
\end{align}
where $\Pi$ is the conjugate momentum of $\xi$ and \(n^\mu = (e^{-\alpha},0,0,0)\) is the 4-velocity of the Eulerian observer.

In terms of \(\Pi\) and \(\Theta\), the equations of motion can be rewritten as a system of partial differential equations suitable for the numerical evolution:
\begin{align}
    \partial_t\xi &= e^{\alpha} \Pi,\label{eq:dt_xi} \\
    \partial_t\Theta &= e^{\alpha}\partial_r \alpha \Pi +e^{\alpha} \partial_r\Pi,\label{eq:dt_th}\\
    \partial_t\Pi &= \frac{e^{\alpha} \left(e^{-2 \beta} \left(r \partial_r\Theta+\Theta \right)+\Theta  \left(Cr^2 \Pi ^2+1\right)\right)}{r},\label{eq:dt_Pi}\\
    \partial_t \beta &= -Cr e^{\alpha} \Theta  \Pi,  \label{eq:dt_B}\\
    \partial_r \beta &= -\frac{e^{2 \beta} \left(Cr^2 \Pi ^2+1\right)+Cr^2 \Theta^2-1}{2 r}, \label{eq:dr_B}\\
    \partial_r \alpha &= \frac{e^{2 \beta} \left(1-Cr^2 \Pi ^2\right)-Cr^2 \Theta ^2-1}{2 r}.\label{eq:dr_A}
\end{align}
The first four relations correspond to the evolution equations for the dynamical variables, while the last two represent the constraint equations that must be satisfied at each time step to ensure the consistency of the system. 

In the absence of a scalar field ($\xi=0$), \(\beta\) is stationary and one obtains exactly the Schwarzschild metric
\begin{equation}
    e^{2\alpha}=e^{-2\beta}=1-\frac{2M}{r}\,.
\end{equation}

\subsection{Numerical integration}  \label{sec:NumericalIntegration}
The numerical integration of Eqs.~\eqref{eq:dt_xi}--\eqref{eq:dt_B} is performed using the method of lines~\cite{atkinson2009numerical}: the evolution equations are integrated in time at each grid point, while the spatial derivatives on the right-hand sides are evaluated by finite differences. Specifically, we employ the fourth order accurate Runge-Kutta method for time integration, and the fourth order accurate centered finite differences formula for the spatial derivatives.
We use a uniform numerical grid, staggered by half a grid step in order to avoid placing a grid point on the origin, where the right-hand sides of Eqs.~\eqref{eq:dt_Pi},~\eqref{eq:dr_B} and~\eqref{eq:dr_A} cannot be evaluated numerically. In other words, the grid points are located at
\begin{equation}
r_j=\left(j+\tfrac12\right)\Delta r,\qquad j=0,\dots,N_p - 1,
\end{equation}
where $\Delta r$ is the grid spacing and $N_p$ the number of grid points.
To apply centered stencils at the first nodes we introduce three ghost zones on the left, at $r_{-1}=-\Delta r/2$, $r_{-2}=-3\Delta r/2$, and $r_{-3}=-5\Delta r/2$: in these cells the variables are not evolved, but at each time step they are filled by enforcing regularity and spherical symmetry. In particular, for the scalars $\alpha$, $\beta$, $\xi$ and $\Pi$ we impose even parity, whereas for $\Theta=\partial_r\xi$ we impose odd parity. At the outer boundary $r_\infty$, we add three ghost zones beyond the last physical grid point, whose values remain frozen after initialization. This is legitimate because the initial profile of $\xi$ is localized close to the center of the spatial domain and rapidly decays toward the outer boundary. Moreover, we choose a combination of the size of the grid and the total time of integration that prevent signals from reaching it. 

The scalar field characteristics have speeds~\cite{Ripley:2019irj}
\begin{equation}
v_{\pm}=\frac{dr}{dt}=\pm e^{\alpha-\beta},
\end{equation}
which tend to $1$ as $r\to\infty$, since $\alpha,\beta\to0$.
Consequently, as long as $t/r_\infty \lesssim 1$, no signal reaches the outer boundary, and the fields there remain constant.
To set the time integration step $\Delta t$, it is convenient to introduce the Courant-Friedrichs-Lewy (CFL) factor
\begin{equation}
\overline{\rm CFL} = v_{\text{char}} \frac{\Delta t}{\Delta r},
\label{eq:CFLdef}
\end{equation}
where $v_{\text{char}}=|v_\pm|$ denotes the propagation speed of signals within the system. Stability of the simulation requires the CFL factor above to be smaller than an upper value which depends on the integration scheme. For the simulations performed in this work, we first choose a fiducial value of the CFL factor, and then set the time step as
\begin{equation}
\Delta t = {\text{CFL}} \, \Delta r.
\label{eq:CFLtimestep}
\end{equation}
The factor ${\text{CFL}}$ corresponds to $\overline{\rm CFL}$ in Eq.~\eqref{eq:CFLdef} when $v_\text{char} \approx 1$. This approximation works well in cases where the spacetime is nearly flat, and the scalar field assumes only small values. However, we will see that, due to our Schwarzschild-like slicing, in our simulations the coordinate characteristic speed $v_{\text{char}}$ can become significantly larger than unity when the initial scalar wave packet has a large amplitude, which requires choosing a sufficiently small value of ${\rm CFL}$.
For each time step, the profile of the metric function $\alpha$ is obtained by integrating the constraint~\eqref{eq:dr_A} along the radial direction. In particular, since Eq.~\eqref{eq:dr_A} has the form
\begin{equation}
    \partial_r \alpha = F[r,\beta,\Theta,\Pi],
\end{equation}
we can obtain the profile of $\alpha$ by integrating the right-hand side as
\begin{equation}
    \alpha(r) = \alpha(r_{\infty+2}) + \int_{r_{\infty+2}}^{r} F\,dr, 
\end{equation}
where $r_{\infty+2}$ is the second ghost zone at the outer boundary. The integral is evaluated with the Simpson's rule, and starting from $r_{\infty+2}$ (where the values of the fields are given by our boundary conditions) allows us to reach fourth order convergence on the last grid point $r_{\infty}$.  

To ensure the stability of the evolution and suppress numerical instabilities arising from high-frequency modes, we introduce a fifth-order Kreiss-Oliger dissipation term~\cite{Kreiss1973MethodsFT,Babiuc:2007vr}, defined as 
\begin{equation}
    \partial_t u \to \partial_t u - \frac{\eta_{\rm KO}}{64\Delta t}(\Delta r)^6(D_+^3 D_-^3 u), \label{eq:KO}
\end{equation}
where $u$ denotes a generic field, $\eta_{\rm KO}$ is a coefficient to be fixed below, while $D_\pm$ are first-order finite difference operators with one-sided backward ($-$) and forward ($+$) schemes. The computation of this correction requires the use of three grid points on each side of the point where it is evaluated; the use of three ghost zones allows us to perform this computation even at the boundaries.

\subsection{Initial data}  \label{IC}

\begin{figure}
    \centering
    \includegraphics[width=\linewidth]{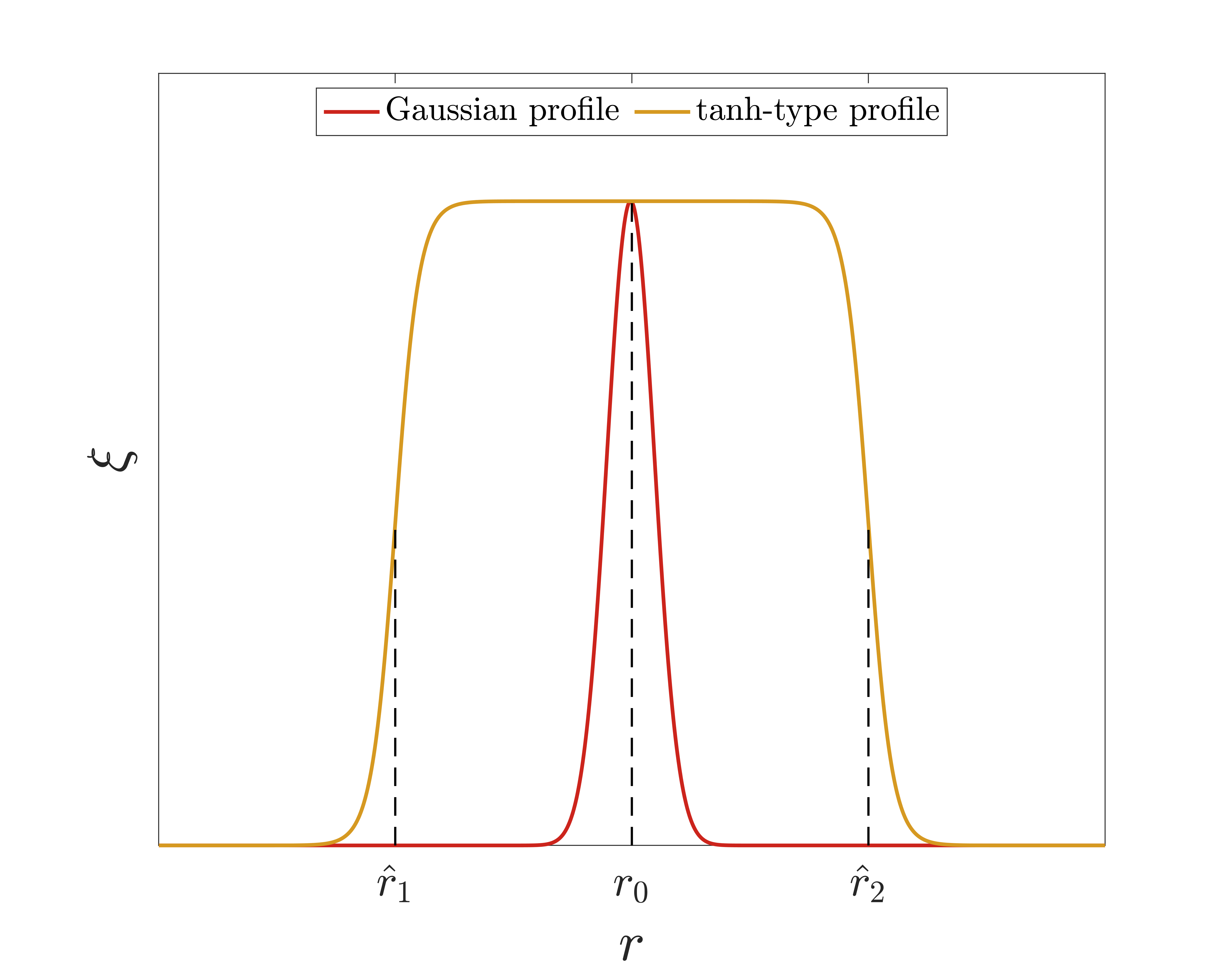}
    \caption{Representative initial profiles of the scalar field $\xi(r)$ used in the numerical simulations. The Gaussian profile (red curve) corresponds to a localized pulse, while the tanh-type profile (yellow curve) produces a smoothed step with an approximately constant plateau.
    }
    \label{fig:init profili}
\end{figure}

As initial data of the scalar field we considered both a Gaussian and a tanh-type profile
\begin{align}
      &\xi(t=0,r)_{\rm Gaussian}
      = \frac{A}{r}\exp\!\left[-\frac{(r-r_0)^2}{2\sigma^2}\right],\label{gauss pacchetto collasso}\\
    &\xi(t=0,r)_{\rm tanh}=\frac{A}{2} \left[ \tanh\!\left(\frac{r-{\hat r}_1}{\sigma}\right) - \tanh\!\left(\frac{r-{\hat r}_2}{\sigma}\right) \right] \label{tanh xi}.
\end{align}
In Eq.~\eqref{gauss pacchetto collasso}, $A$ denotes the field amplitude, $r_0$ the initial central position of the pulse, and $\sigma$ its characteristic width. 
In Eq.~\eqref{tanh xi}, the parameters ${\hat r}_1$ and ${\hat r}_2$ define the left and right edges of the tanh-like profile, respectively, and thus determine the radial interval over which the function $\xi(r)$ attains a value close to the plateau $A$. More precisely, for $r \lesssim {\hat r}_1$ and $r \gtrsim {\hat r}_2$ the function rapidly approaches zero, whereas in the intermediate region ${\hat r}_1 \lesssim r \lesssim {\hat r}_2$ it remains approximately constant (see Fig.~\ref{fig:init profili} for a schematic representation). The separation ${\hat r}_2-{\hat r}_1$ therefore sets the effective width of the smoothed step, while the parameter $\sigma$ controls the sharpness of the transition between the near-vanishing regions and the central plateau.

After setting the profile of $\xi$, the auxiliary fields $\Theta$ and $\Pi$ are initialized as
\begin{align}
\Theta(t=0,r)&=\partial_r\xi(t=0,r),\label{ingoing1}\\
\Pi(t=0,r)&=\xi(t=0,r)/r+\Theta(t=0,r)\label{ingoing2}.     
\end{align}
The choice in Eq.~\eqref{ingoing2} ensures that the initial wave is predominantly ingoing. The initial data are chosen so that the main part of the wave packet is initially concentrated near the inner boundary. The profiles of the metric functions $\alpha(t=0, r)$ and $\beta(t=0, r)$ are obtained by integrating numerically the constraints~\eqref{eq:dr_A} and~\eqref{eq:dr_B} starting from the first grid point $r_1=\Delta r/2$ and moving outward. In order to obtain the initial values for the integration procedure, we perform a Taylor expansion of the metric function at the origin
\begin{align}
    \alpha(t,r) &= \alpha_0(t) + \alpha_2(t)\,r^2 + \mathcal{O}(r^4),  \label{eq:TaylorAOrigin} \\
    \beta(t,r) &= \beta_0(t) + \beta_2(t)\,r^2 + \mathcal{O}(r^4),  \label{eq:TaylorBOrigin}
\end{align}
in which terms of odd power in $r$ vanish, consistently with parity. Regularity of the metric at $r = 0$ requires that $e^{2\beta(t,0)} = 1$, and hence $\beta_0 = 0$. $\alpha_0$ is instead a free parameter as, due to our choice of the metric ansatz~\eqref{eq:metric}, a shift of the profile of $\alpha$ can be compensated with a rescaling of the coordinate time. We therefore decide to initially set $\alpha_0 = 0$. Substituting the expansions~\eqref{eq:TaylorAOrigin} and~\eqref{eq:TaylorBOrigin} in the constraint equations we can compute $\alpha_2$ and $\beta_2$. The values of the metric functions at the first grid point $r_1$ then read
\begin{align}
    \alpha(t=0,r_1) &= -\frac{C}{3}\,\Pi^2(r_1)\,r_1^2, \\
    \beta(t=0,r_1) &= -\frac{C}{6}\,\Pi^2(r_1)\,r_1^2.
\end{align}
After performing the numerical integration of Eqs.~\eqref{eq:dr_A} and~\eqref{eq:dr_B} with the fourth order accurate Runge-Kutta method up to the outer boundary, we shifted the profile of $\alpha(t = 0, r)$ in order to impose
\begin{equation}
    \alpha(r_\infty) = 0,
\end{equation}
which corresponds to requiring that the coordinate time $t$ coincides with the proper time at spatial infinity.

\begin{figure}
    \centering
    \includegraphics[width=\linewidth]{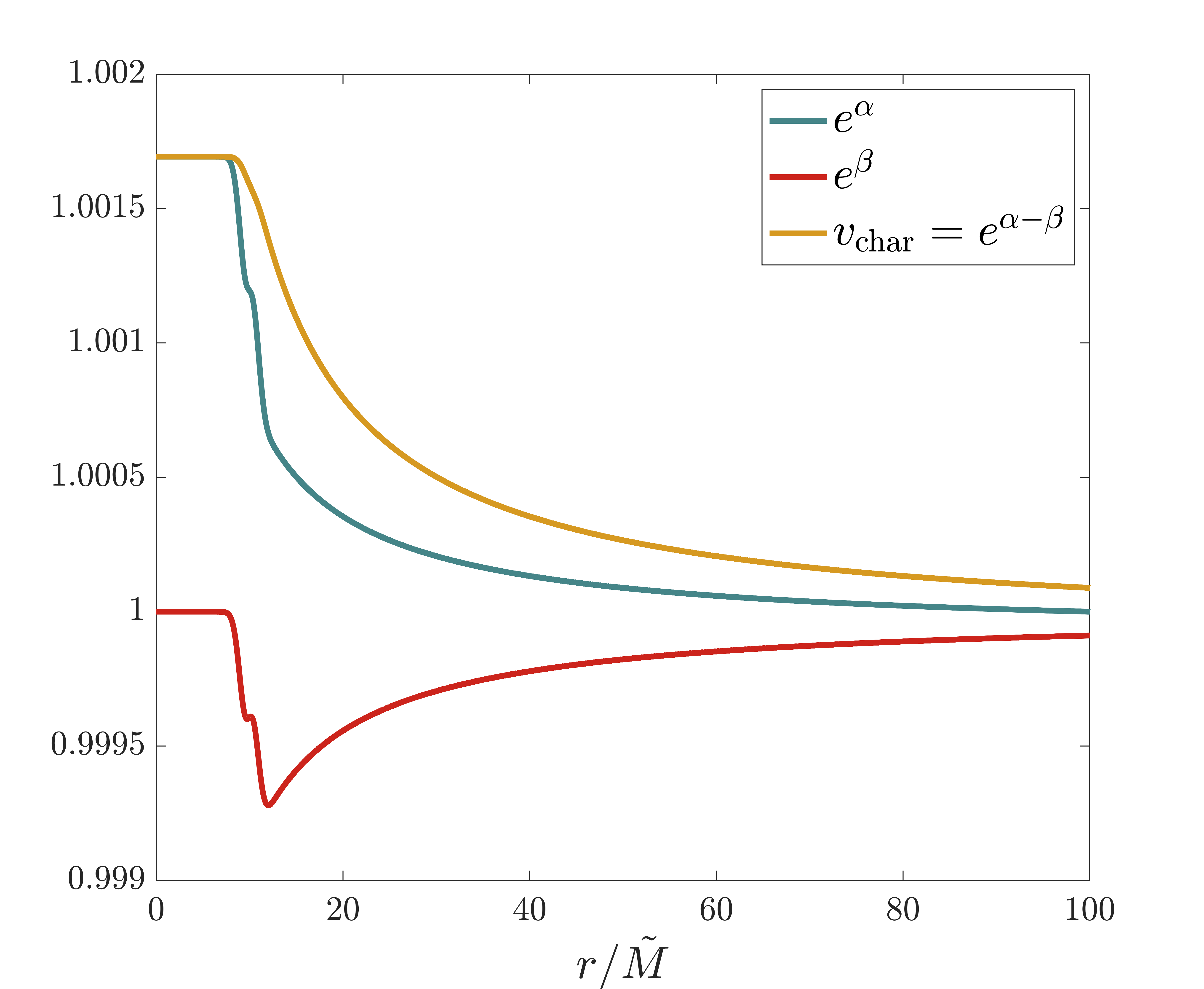}
    \caption{Initial profiles for $e^\alpha$, $e^\beta$ and $v_{\rm char}$, describing a nearly flat spacetime locally perturbed by a Gaussian pulse of the phantom field centered at $r_0=10\tilde{M}$. The amplitude of the wave packet is $A=0.1\tilde{M}$ and its width is $\sigma=\tilde{M}$.}
    \label{fig:confronto}
\end{figure}

In Fig.~\ref{fig:confronto} we show, as an example, the initial profiles of $e^\alpha$, $e^\beta$ and $v_{\rm char}$ corresponding to a Gaussian wave packet centered in $r_0 = 10\tilde{M}$, having initial amplitude $A=0.1\tilde{M}$ and width $\sigma=\tilde{M}$. As we can see the metric is essentially flat for values of $r$ far from the wave position and inside it, while around $r \approx 10\tilde{M}$ both $e^\alpha$ and $e^\beta$ decrease due to the presence of the Gaussian pulse. After that, the metric functions start approaching unity, consistently with asymptotic flatness.

\subsection{Diagnostic variables}
\label{diagnostic collasso}

To investigate the numerical evolution of the phantom-field collapse, we introduce a set of diagnostic quantities to monitor the geometric and physical properties of the spacetime. Among these, a central role is played by the Kretschmann scalar, which provides a coordinate-invariant measure of the spacetime curvature~\cite{Henry:1999rm}. It is defined as
\begin{align}
K = R_{\mu\nu\rho\sigma} R^{\mu\nu\rho\sigma} &=\frac{8 C r^2 e^{-4 \beta}}{r^4}\left(\Theta^2-e^{2 \beta}\Pi^2\right)\notag\\
&\;\times\left(-e^{2 \beta}\left(C r^2 \Pi^2+1\right)+C r^2 \Theta^2+1\right)\notag\\
&\;+\frac{12 e^{-4 \beta}}{r^4}\left(e^{2 \beta}-1\right)^2.
\label{Krest}
\end{align}
where $R_{\mu\nu\rho\sigma}$ is the Riemann tensor. 
The above expression was obtained by explicitly computing the Riemann tensor $R_{\mu\nu\rho\sigma}$ from the line element~\eqref{eq:metric} and subsequently using the evolution system~\eqref{eq:dt_xi}-\eqref{eq:dr_A} to express the result in terms only of the dynamical variables. Inspecting the spacetime behavior of $K$ enables us to assess whether singularities develop during the simulations. In particular, we monitor whether $K$ remains finite and regular throughout the evolution or exhibits divergent behavior in localized regions, which would signal the formation of curvature singularities.

Alongside the Kretschmann scalar, we monitor the Misner-Sharp mass~\cite{Abreu:2010ru}, which in our coordinates takes the form
\begin{equation}
    M_{\rm MS}(t,r) = \frac{r}{2} \left( 1 - g^{\mu\nu}\,\partial_\mu r\,\partial_\nu r \right)
    = \frac{r}{2}\left(1 - e^{-2\beta(t,r)}\right).
\end{equation}
This quantity provides a quasi-local measure of the gravitational mass contained within a sphere of areal radius $r$. In spherical symmetry, it also serves as a convenient diagnostic for the formation of apparent horizons. Indeed, trapped surfaces are characterized by the condition
\begin{equation}
\label{condizione mmms}
    \frac{2M_{\rm MS}(t,r)}{r} \geq 1,
\end{equation}
which signals the vanishing of the expansion of future directed outgoing null geodesics.

In the phantom-field case, however, the effective energy density can be negative, contributing negatively to the mass function. One therefore expects that Eq.~\eqref{condizione mmms} is not satisfied, thereby preventing the formation of trapped regions.

In the case of a canonical scalar field ($C=-1$), one instead expects the formation of an apparent horizon, provided the amplitude of the initial wave packet is sufficiently large. However, within the metric ansatz in Eq.~\eqref{eq:metric}, the coordinates are not horizon-penetrating. As a result, black-hole formation cannot be reliably identified solely by tracking the regions where condition~\eqref{condizione mmms} is satisfied, since coordinate effects may obscure the approach to horizon formation.

Following Ref.~\cite{Ripley:2019irj}, we therefore monitor the characteristic speeds of the system, which encode information about its causal structure. In particular, apparent horizons can be identified as the locations where the characteristic speed of outgoing radial null rays, $v_{+}$, vanishes. Physically, this corresponds to the point where outward-propagating light fronts become momentarily frozen, signaling the onset of a trapped region.
Operationally, we define the apparent horizon radius as the outermost spherical shell where $v_{+}$ drops below a prescribed threshold $\varepsilon$~\cite{Ripley:2019irj,Thaalba:2023fmq}:
\begin{equation}
\label{criterio epsilon rah}
    r_\text{AH} = \max \left\{ r \;\middle|\; 
    v_+ = \exp\!\big(\alpha(t,r)-\beta(t,r)\big) < \varepsilon \right\}.
\end{equation}
We will typically take $\varepsilon=10^{-6}$ as a fiducial value.

The combined use of the Kretschmann scalar and this horizon-finding criterion allows us to characterize the end state of the spherical collapse, distinguishing between dispersion, black-hole formation, and potential violations of cosmic censorship or the emergence of alternative stationary configurations.

\section{Collapse of wave packets of a canonical scalar field}  \label{sec:resultscanonical}

In order to validate our numerical implementation, in addition to the convergence analysis reported in Appendix~\ref{app:convergence}, we simulated the gravitational collapse of a canonical scalar field and verified that we recover the key features of Choptuik's seminal results~\cite{Choptuik:1992jv}. In particular, we checked for the existence of a critical amplitude $A^*$, independent of numerical parameters, that marks the threshold between dispersion and black-hole formation.

To this end, we set $C=-1$ and considered Gaussian initial data with fixed parameters $\sigma=\tilde{M}$ and $r_0=10\tilde{M}$, varying only the amplitude $A$ of the scalar field.

\begin{figure*}
    \centering
    \includegraphics[width=\columnwidth]{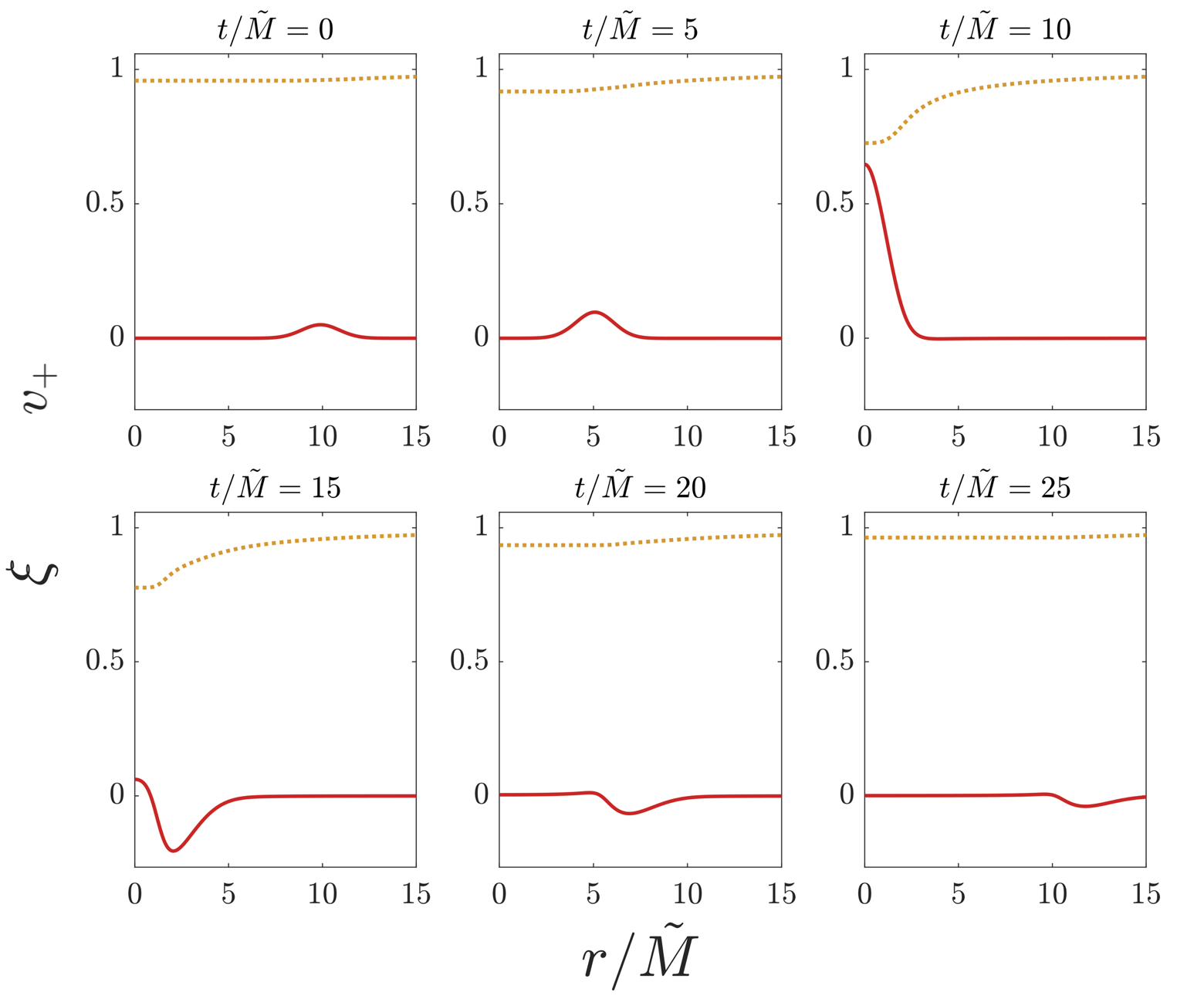}
    \includegraphics[width=\columnwidth]{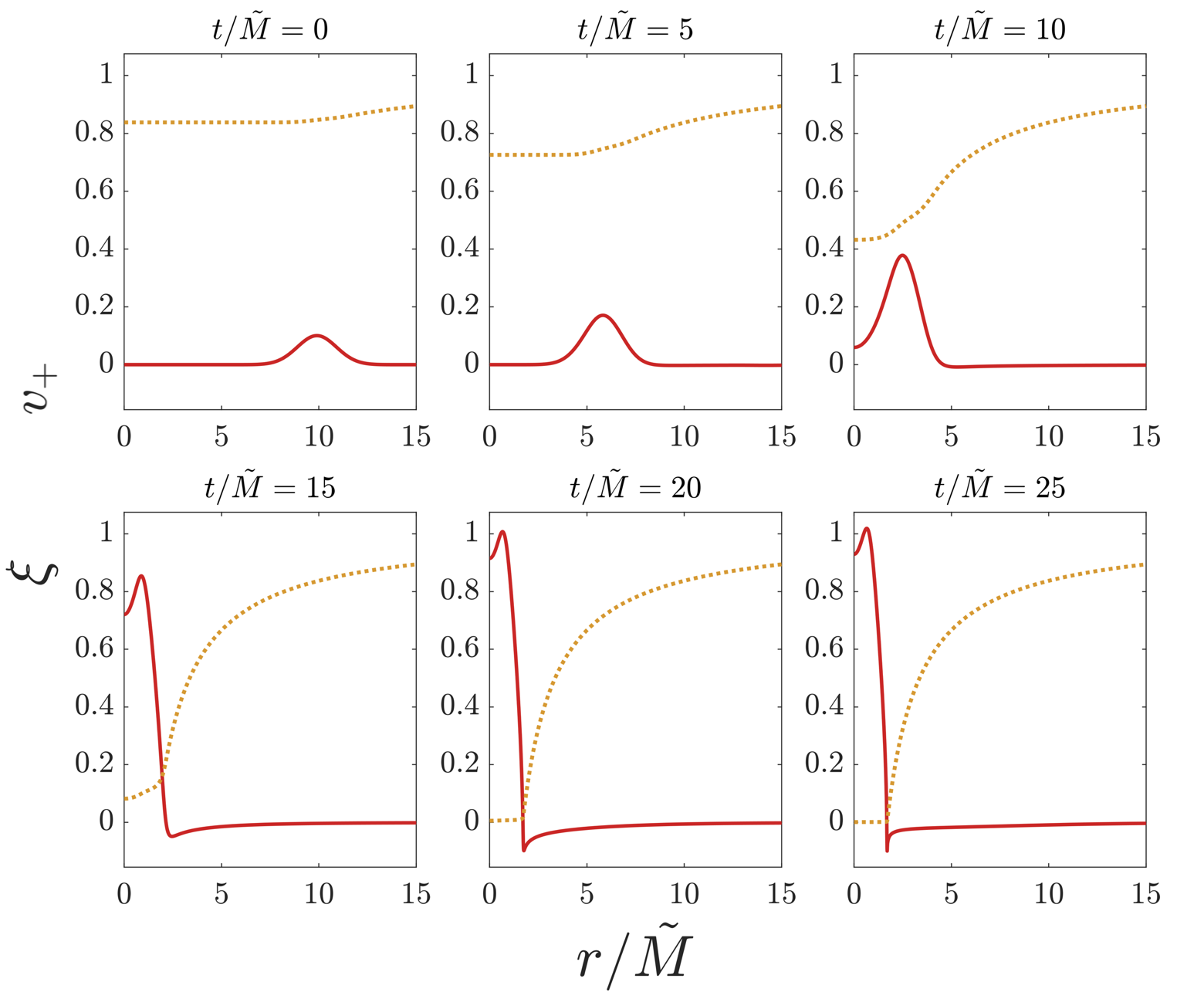}
    \caption{
    Nonlinear evolution of a standard scalar field (red solid curve) and of the outgoing characteristic velocity $v_{+}$ (yellow dashed curve) for a Gaussian initial profile with $\sigma=\tilde{M}$, $r_0=10\tilde{M}$. 
    Each small panel corresponds to a different time snapshot. 
    Left panels: Initial amplitude $A=0.2\tilde{M}$; the field disperses and no horizon forms.
    Right panels: Initial amplitude $A=\tilde{M}$; the vanishing of $v_{+}$ within a finite region signals the formation of a trapped surface and an apparent horizon.}
    \label{fig:Choptuik}
\end{figure*}

For sufficiently small amplitudes, the scalar field disperses, leaving behind a regular spacetime. An illustrative example is shown in the left panels of Fig.~\ref{fig:Choptuik}, corresponding to $A=0.2\tilde{M}$. Each panel displays a different time snapshot of the evolution. The red solid curve represents the scalar-field profile, while the yellow dashed curve shows the outgoing characteristic velocity $v_{+}$. The wave packet initially propagates inward ($t \lesssim 10\tilde{M}$), reaches the central region, and subsequently disperses ($t \gtrsim 10\tilde{M}$). Throughout the evolution, the characteristic velocity remains close to unity and never approaches zero, indicating that no trapped region forms.

For larger initial amplitudes, the evolution instead leads to black-hole formation. This behavior is illustrated in the right panels of Fig.~\ref{fig:Choptuik}, which show the case $A=\tilde{M}$ using the same conventions as in the left panels. As time progresses, the outgoing characteristic velocity $v_{+}$ decreases and eventually approaches zero within a finite radial interval. In this region, the evolution effectively freezes, reflecting the fact that outgoing light rays can no longer propagate to larger radii. This signals the formation of a trapped region, which we identify as an apparent horizon.

Using the criterion of Eq.~\eqref{criterio epsilon rah}, we estimate the apparent-horizon radius to be $r_{\text{AH}} = 1.653\tilde{M}$.
At the time of formation, the Misner-Sharp mass extracted at the outermost point of the physical grid is $M_{\rm MS} = 0.849\tilde{M}$, in good agreement with the Schwarzschild relation $r_{\text{H}}=2M$. The small deviation from exact equality is expected, since the initial data are not purely ingoing and residual scalar-field energy remains outside the horizon. These results provide clear numerical evidence of black-hole formation within the chosen configuration.

Following Choptuik's analysis, we systematically varied the amplitude $A$ while keeping $\sigma$ and $r_0$ fixed, and identified the critical value $A^*$ separating dispersive evolutions from black-hole formation. To ensure robustness, this study was repeated at different spatial resolutions and CFL factors.

\begin{figure}
    \centering
    \includegraphics[width=\linewidth]{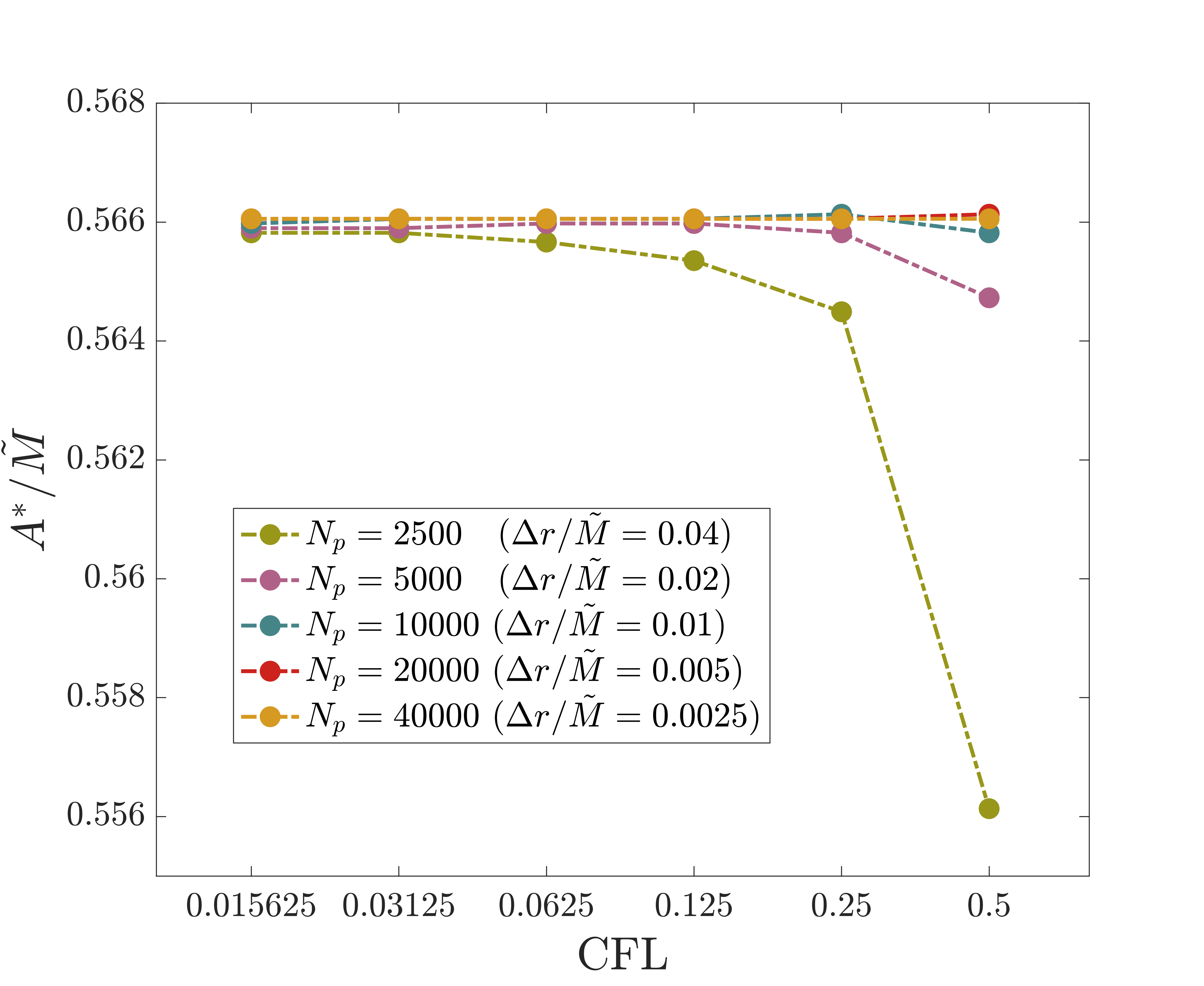}
    \caption
    {Critical amplitude $A^*$ separating dispersion from black-hole formation as a function of spatial resolution $N_p$. Different colors denote different CFL factors. As the resolution increases, $A^*$ converges toward $0.566\tilde{M}$, demonstrating numerical robustness.}
    \label{fig:convergenza standard}
\end{figure}

The results are shown in Fig.~\ref{fig:convergenza standard}, where we plot the critical amplitude $A^*$ as a function of the CFL factor, with different colors corresponding to different spatial resolutions (labeled by the number of grid points $N_p$). As the resolution increases or the CFL factor decreases, the critical amplitude converges toward
\[
A^* \simeq 0.566\tilde{M}.
\]
The stabilization of $A^*$ at high resolution confirms that the code reliably captures the threshold between dispersion and collapse, and reproduces the expected critical behavior of canonical scalar-field collapse.

\section{Collapse of wave packets of a phantom scalar field}  \label{sec:resultsphantom}

We now turn to the case of a phantom scalar field, corresponding to $C = +1$. All numerical simulations are performed on a radial domain $r \in [0,100]\tilde{M}$. Unless otherwise specified, the coefficient of the Kreiss-Oliger dissipation operator in Eq.~\eqref{eq:KO} is fixed to $\eta_{\rm KO}=0.1$.

\subsection{Breakdown and behavior of the phantom field}

From preliminary runs, at fixed numerical parameters, we observe that for sufficiently small amplitudes the wave packet disperses, similarly to the canonical case ($C = -1$). However, for larger values of the amplitude $A$, the numerical evolution eventually breaks down. While this behavior could in principle be purely numerical, it may also signal a qualitative transition in the underlying dynamical regime.
 
To investigate the origin of this behavior, we perform a set of simulations aimed at identifying the threshold amplitude, $A_\text{break}$, above which the numerical evolution breaks down. We begin by considering Gaussian initial data with $\sigma=\tilde{M}$ and $r_0=10\tilde{M}$, and determine $A_\text{break}$ through a bisection procedure, iterated until a target accuracy $\varepsilon_A = 10^{-4}$ is reached. We then repeat the analysis while varying the numerical parameters, including the grid spacing, the time step, and the Kreiss-Oliger dissipation coefficient $\eta_{\rm KO}$. The underlying rationale of our investigation is the following: if $A_\text{break}$ converges toward a stable value under grid refinement and variations of numerical parameters, the breakdown can be attributed to a physical mechanism intrinsic to the dynamics. Conversely, a strong sensitivity of $A_\text{break}$ to resolution would indicate that the observed behavior is primarily of numerical origin.

\begin{figure}
    \centering
    \includegraphics[width=\linewidth]{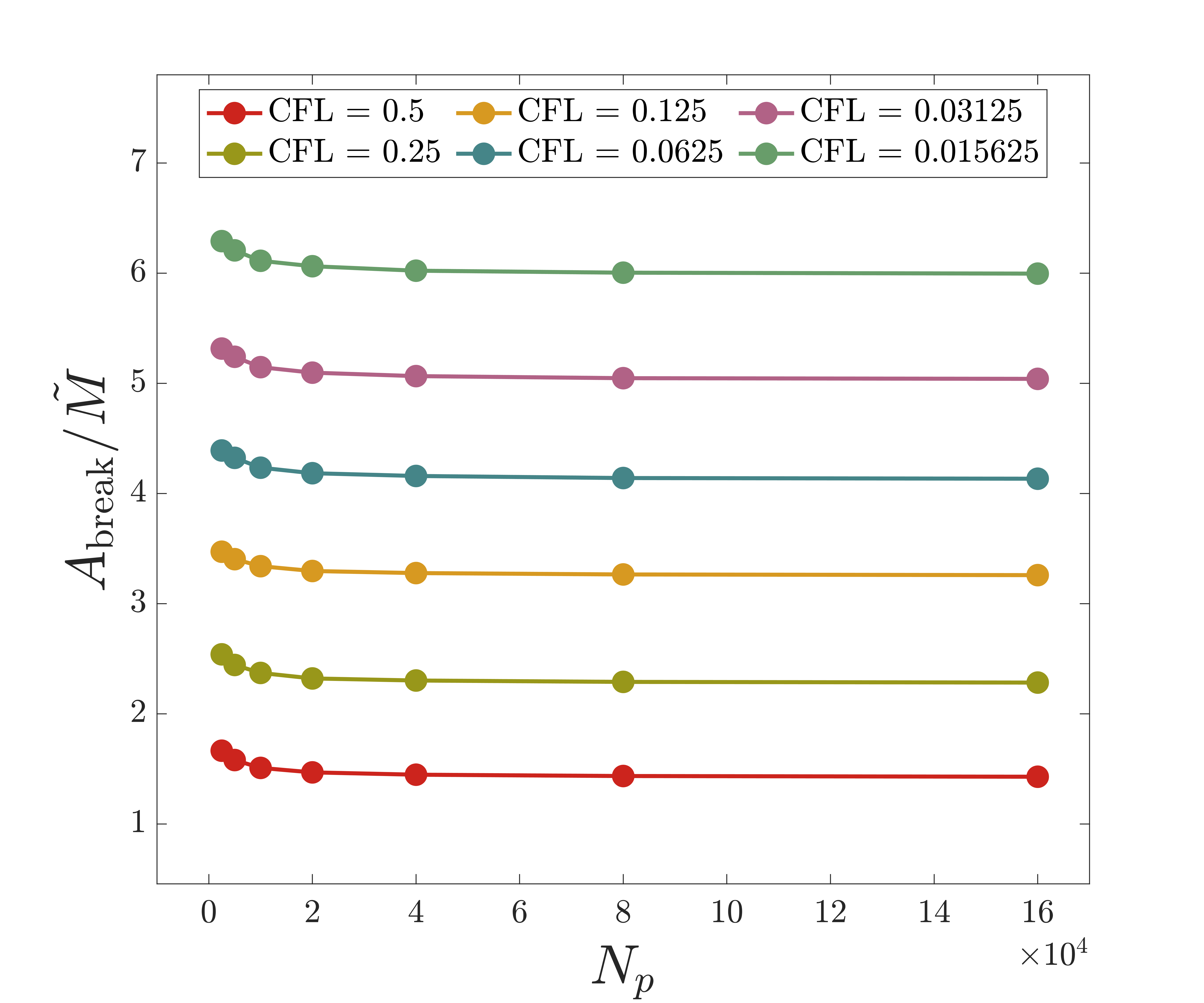}
    \caption{Critical threshold $A_\text{break}$ as a function of the number of radial grid points $N_p$ for different values of the CFL factor. The curves do not converge toward a common continuum limit; instead, $A_\text{break}$ systematically increases as the CFL factor is reduced. This behavior indicates that the observed breakdown is of numerical origin and that the resolution of the dynamics is primarily controlled by the choice of time step.}
    \label{fig:Astar_N}
\end{figure}

The results of our analysis are summarized in Fig.~\ref{fig:Astar_N}, where each curve corresponds to a fixed value of the CFL factor and shows the dependence of $A_\text{break}$ on the number of radial grid points $N_p$. Here we focus only on the dependence on these two numerical parameters; the effect of the Kreiss-Oliger dissipation coefficient, $\eta_{\rm KO}$, is subleading and is discussed in Appendix~\ref{app:KreissOligerImpact}. For each fixed value of the CFL factor, the convergence pattern with increasing spatial resolution is similar: the sequence $A_\text{break}(N_p)$ converges monotonically toward an asymptotic value. At fixed spatial resolution, however, the critical amplitude $A_\text{break}$ systematically increases as the CFL factor is reduced. In other words, decreasing the time step while keeping $\Delta r$ fixed shifts the threshold for numerical breakdown toward larger values of $A$.

The absence of a common asymptotic curve toward which the different sequences converge indicates that the observed breakdown is of numerical origin. Indeed, within the range of amplitudes considered, a sufficiently small CFL factor restores a regular evolution. This strongly suggests that the breakdown is primarily controlled by the temporal resolution rather than by a physical transition in the dynamics.

To further characterize the physical behavior of the system and clarify the origin of the numerical breakdown, we examine the spacetime evolution of the Kretschmann scalar for representative configurations. We consider Gaussian initial data with $r_0 = 10\tilde{M}$ and $\sigma = \tilde{M}$, using $N_p = 20000$ radial grid points.

\begin{figure*}
\centering
\includegraphics[width=\columnwidth]{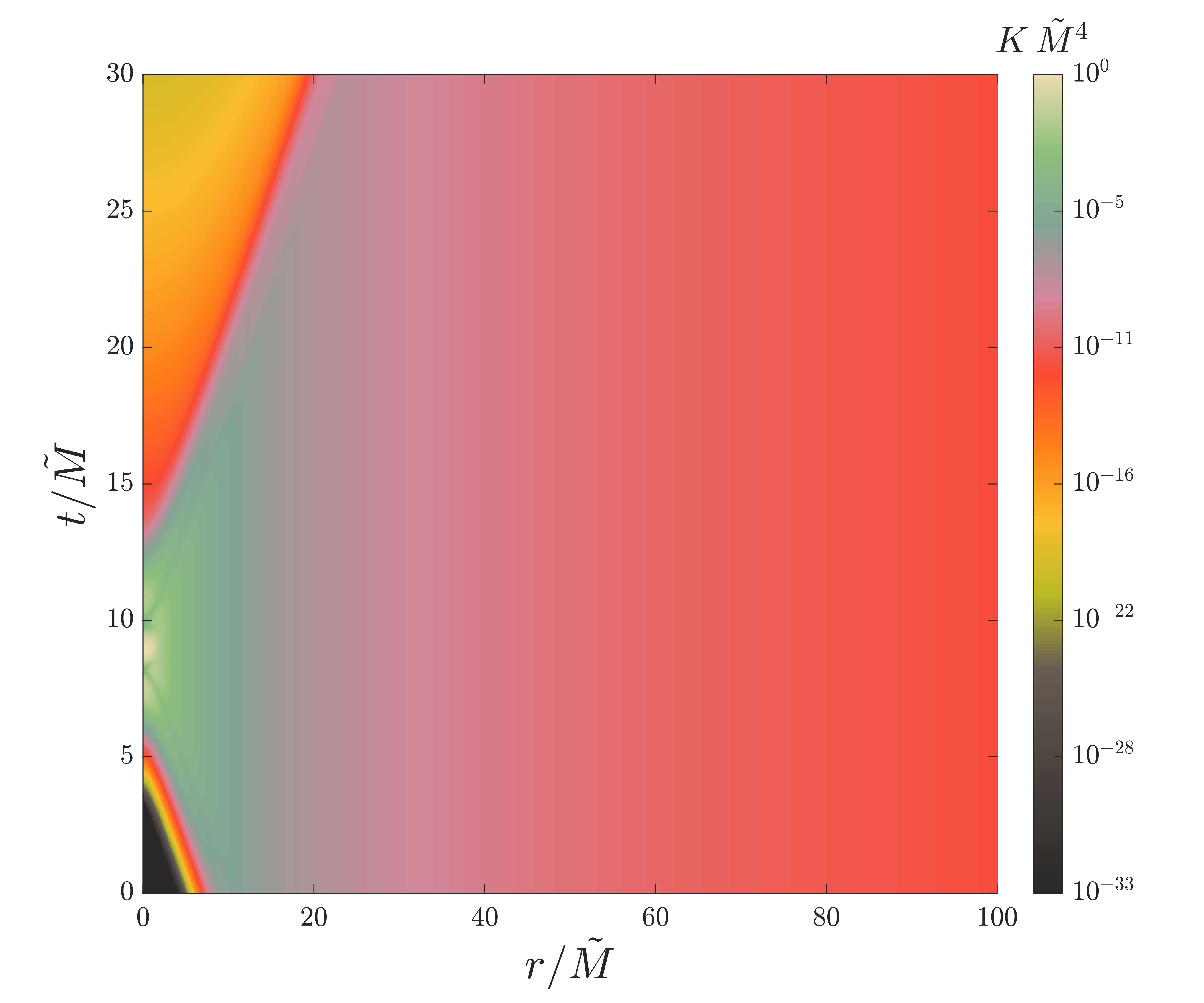} 
\includegraphics[width=\columnwidth]{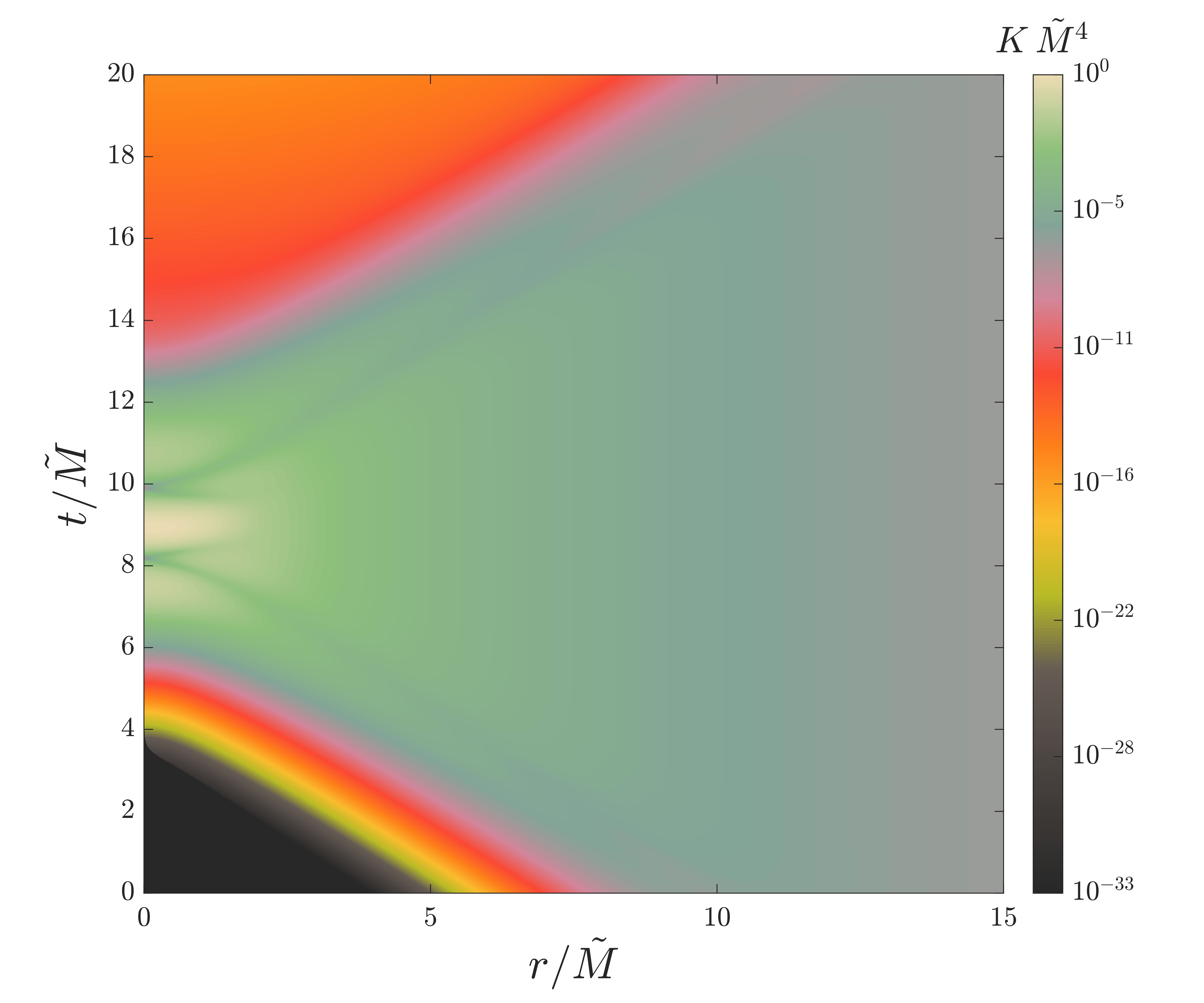}
\caption{%
Spacetime evolution of the Kretschmann scalar $K(t,r)$ for a Gaussian phantom wave packet with $A = 0.5\tilde{M}$, $\sigma=\tilde{M}$, and $r_0 = 10\tilde{M}$. The left panel shows the full domain, while the right panel zooms into the central region.}
\label{fig:A=0.5 colormap}
\end{figure*}

We first examine a simulation with moderate amplitude, $A = 0.5\tilde{M}$, and $\mathrm{CFL} = 0.0625$. 
The left panel of Fig.~\ref{fig:A=0.5 colormap} displays the Kretschmann scalar $K(t,r)$ over the full computational domain ($r \in [0,100]\tilde{M}$, $t \in [0,30]\tilde{M}$), while the right panel shows a zoom into the central region ($r \in [0,12]\tilde{M}$, $t \in [0,20]\tilde{M}$).
The wave packet propagates inward from $r \approx 10\tilde{M}$ and reaches the center at $t \approx 8\tilde{M}$. Around $t \approx 8\tilde{M}\text{--}10\tilde{M}$ the curvature exhibits a transient peak, after which the phantom field disperses outward. In the outer regions, where the scalar amplitude becomes negligible, the geometry asymptotically approaches flat spacetime, albeit modified by the negative effective mass of the configuration. Throughout the evolution the curvature remains finite, and no trapped surfaces form.

\begin{figure}
\centering
\includegraphics[width=\columnwidth]{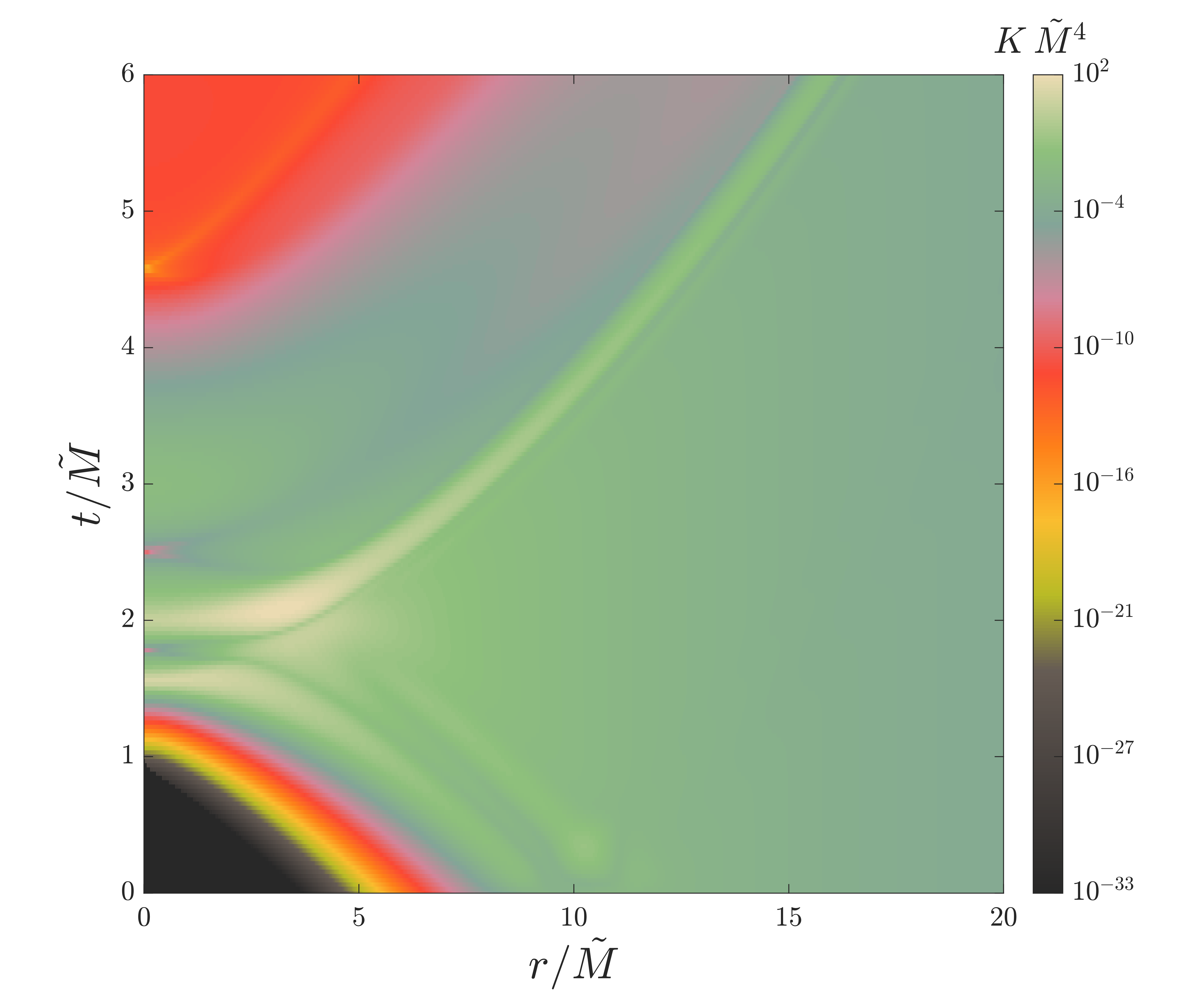}
\caption{%
Spacetime evolution of the Kretschmann scalar $K(t,r)$ for a Gaussian phantom wave packet with amplitude $A= 3.0\tilde{M}$.}
\label{fig:K_A3}
\end{figure}

Increasing the amplitude to $A = 3.0\tilde{M}$ (still with $\mathrm{CFL} = 0.0625$) leads to qualitatively similar dynamics, shown in Fig.~\ref{fig:K_A3} for the central region ($r \in [0,12]\tilde{M}$, $t \in [0,6]\tilde{M}$). The curvature structure becomes more intricate, featuring oblique ridges corresponding to regions of enhanced curvature. The main bright band represents the outgoing front generated after the initial compression phase, while subsequent weaker structures correspond to rarefaction tails and partial reflections. The peak curvature reaches $K \sim 10^2\tilde{M}^{-4}$, about two orders of magnitude larger than in the $A = 0.5\tilde{M}$ case, yet remains finite. No indication of horizon formation or curvature divergence is observed.

Physically, this behavior reflects the repulsive character of the \textit{phantom} field: the negative kinetic term introduces an effective negative energy density that counteracts the initial focusing. The curvature grows rapidly during the initial compression, reaches a maximum, and is then followed by reflection and dispersion rather than collapse.

\begin{figure}
    \centering
    \includegraphics[width=\columnwidth]{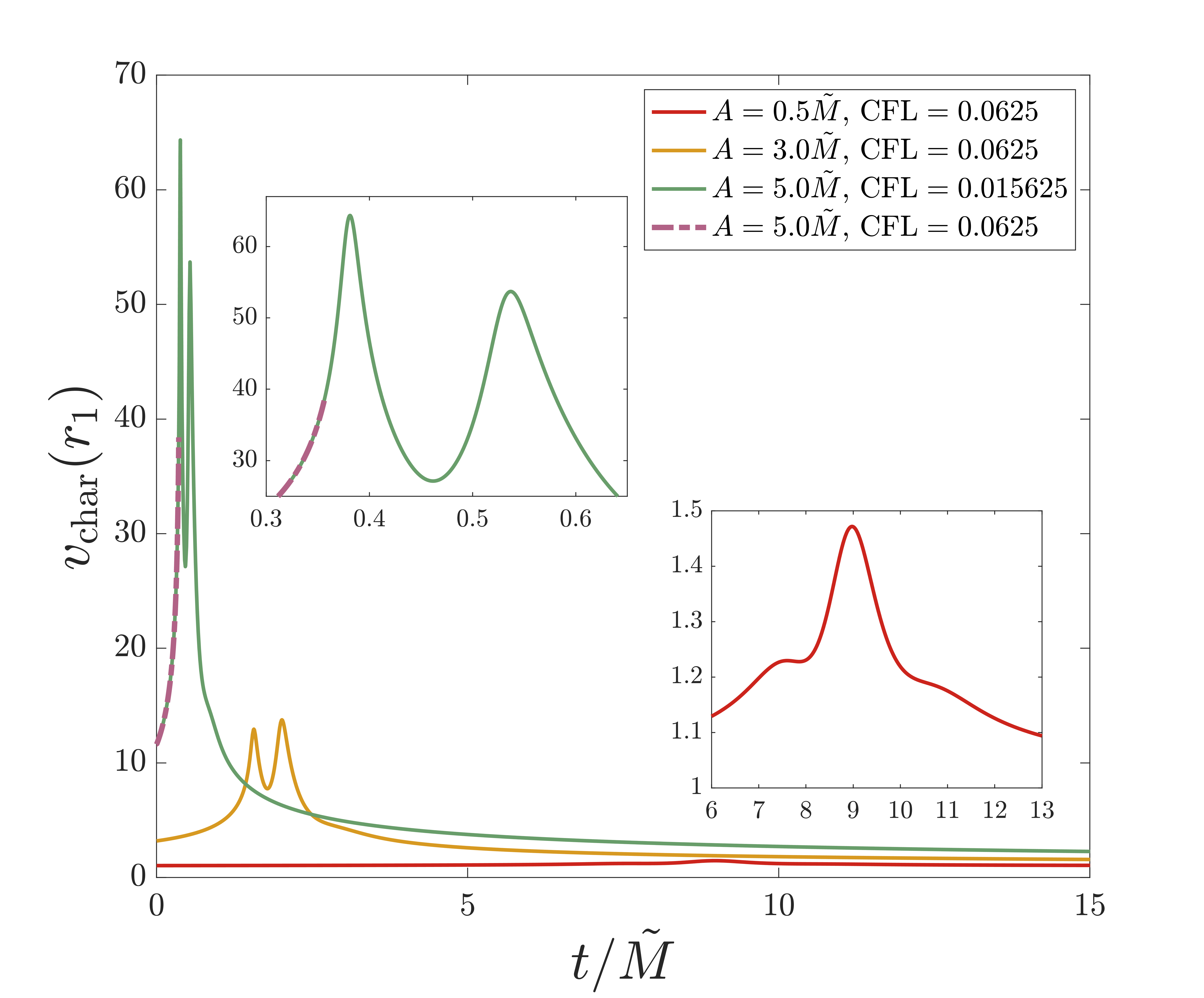}
    \caption{Time evolution of the absolute value of the characteristic velocity, $v_{\rm char}(r_1)$, evaluated at the innermost physical grid point, for Gaussian wave packets with width $\sigma=\tilde{M}$ and center $r_0=10\tilde{M}$. The curves correspond to different amplitudes of the scalar field. The red, yellow and purple curves denote simulations with Courant factor $\mathrm{CFL}=0.0625$, while the green curve corresponds to a smaller value $\mathrm{CFL}=0.015625$ for the simulation with the largest amplitude. The insets show zoomed views of the highlighted region for the low-amplitude case $A=0.5\tilde{M}$, and for the high-amplitude case, $A=5.0\tilde{M}$.}
    \label{fig:vcharconfronto}
\end{figure}

Interestingly, the process now occurs on a much shorter timescale, around $t \simeq 2\tilde{M}$, due to the fact that the larger (negative) effective mass of the phantom field induces larger modifications on metric functions and causes the coordinate characteristic speeds to increase. Such effect is shown in Fig.~\ref{fig:vcharconfronto}, where we plot the evolution of the absolute value of the characteristic velocities, $v_{\rm char}$, extracted at the innermost physical grid point, for the simulations with $A = 0.5\tilde{M}$, $A = 3.0\tilde{M}$ and $A = 5.0\tilde{M}$ (to be discussed below). As we can see, the characteristic speeds grow during the evolution and depend significantly on the initial amplitude, changing by orders of magnitude. Overall $v_{\rm char}$ grows with $A$, causing the dynamics to be faster.

This effect also clarifies the origin of the numerical breakdown and the observed dependence of the critical threshold on the CFL factor.
We operationally defined the CFL factor as in Eq.~\eqref{eq:CFLtimestep}, ${\rm CFL}=\Delta t/\Delta r$, which corresponds to the actual definition~\eqref{eq:CFLdef} only when $v_{\rm char} = 1$.
However, since for large values of $A$ the characteristic speeds increase substantially, the original CFL factor in Eq.~\eqref{eq:CFLdef}, $\overline{\rm CFL}=v_{\rm char}\, {\rm CFL} $, can exceed the upper bound for stability and lead the numerical evolution to a breakdown.
For a given initial amplitude, decreasing ${\rm CFL}$ also reduces $\overline{\rm CFL}$ and can restore stability, postponing the breakdown to larger values of the amplitude and the characteristic velocity.

\begin{figure*}
\centering
\includegraphics[width=\columnwidth]{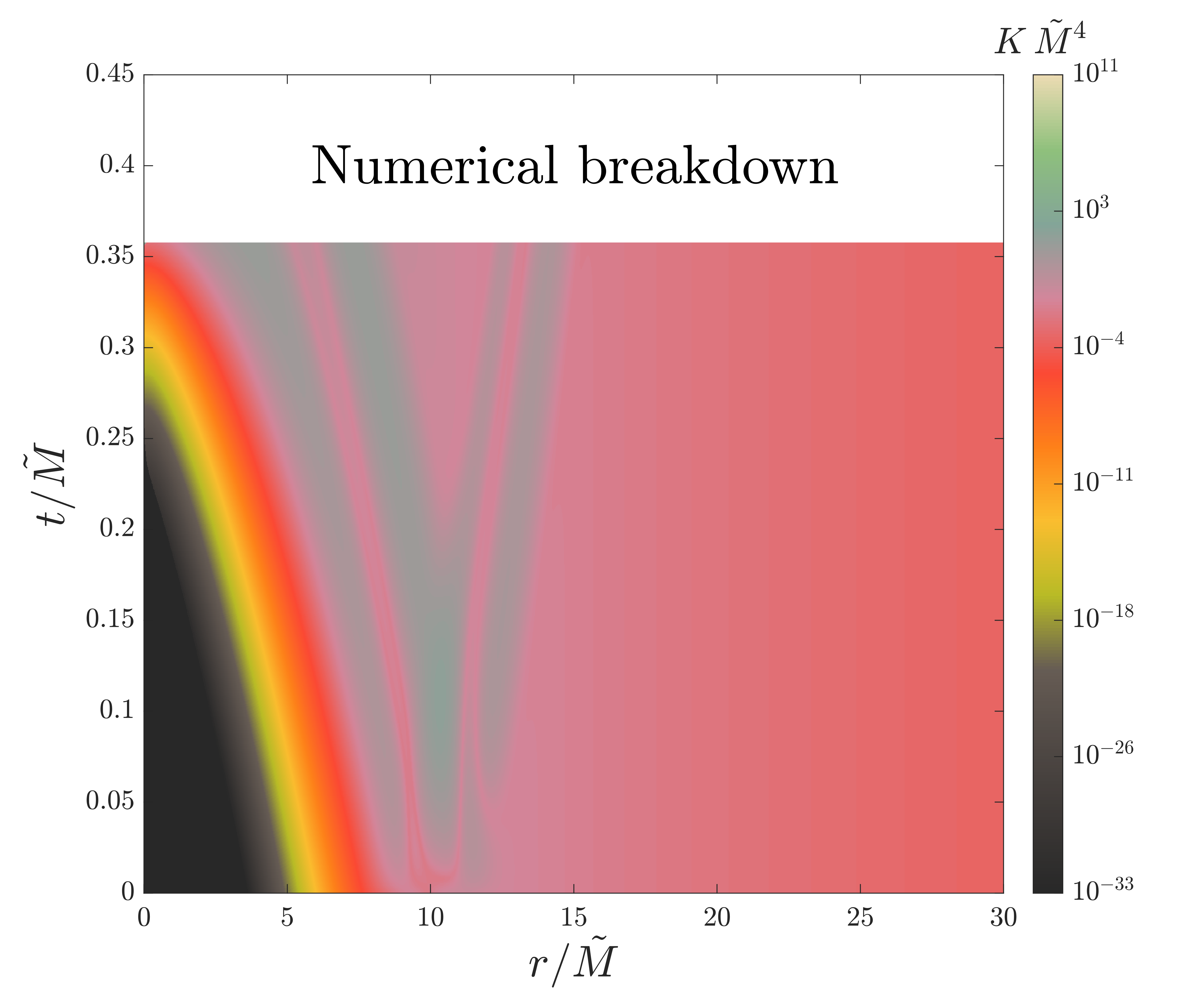}
\includegraphics[width=\columnwidth]{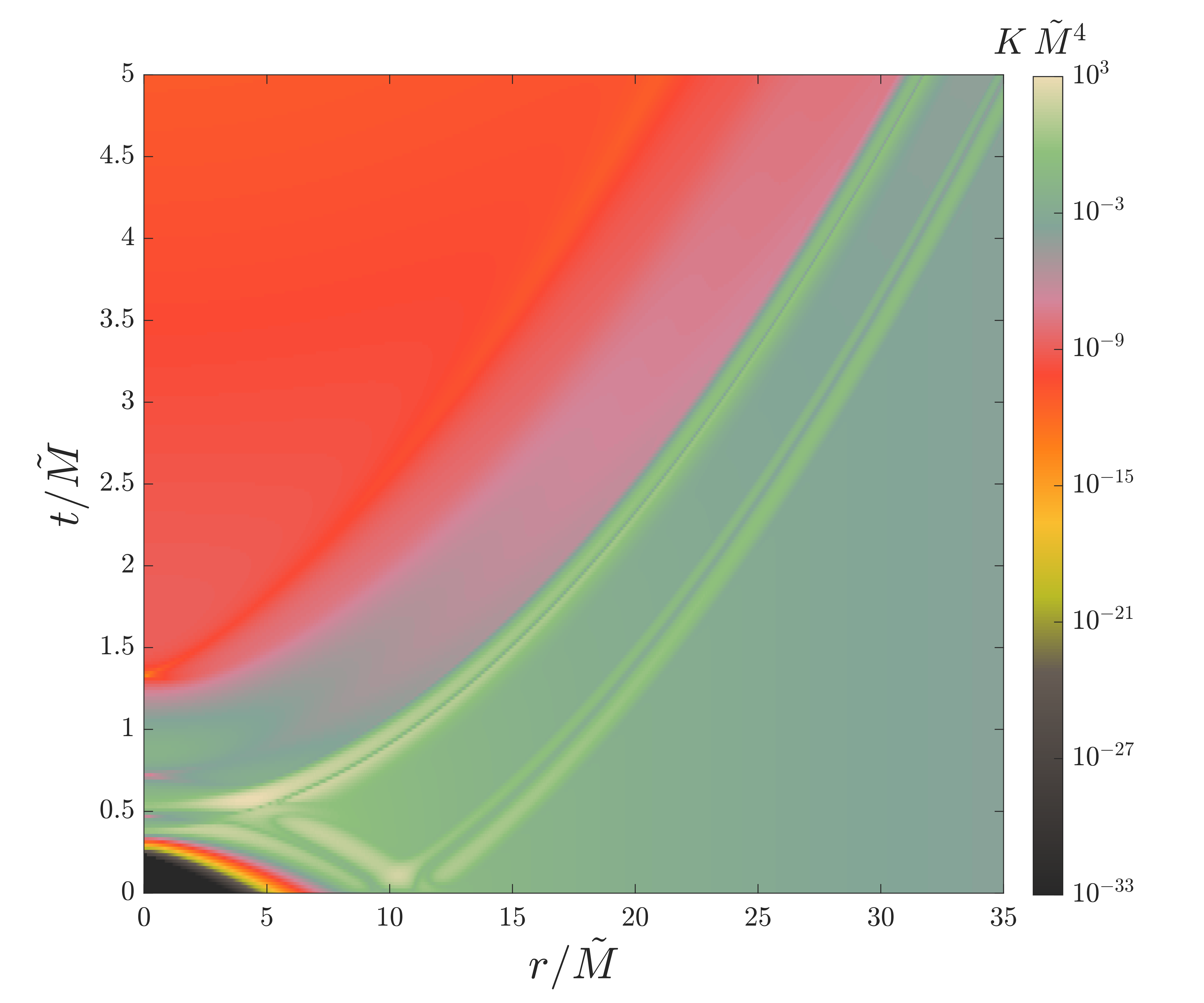}
\caption{%
Left:
Spacetime evolution of the Kretschmann scalar $K(t,r)$ for a Gaussian phantom wave packet with $A = 5\tilde{M}$ and $\mathrm{CFL}=0.0625$. The evolution undergoes numerical breakdown.
Right: Same configuration as in the left panel, but with $\mathrm{CFL}=0.015625$. The evolution remains regular.
}
\label{fig:K_A5_highCFL_breakNobreak}
\end{figure*}

Lastly, we study the case of a very large initial amplitude, $A= 5\tilde{M}$, for which nonlinear effects become even more pronounced. 
In this regime the characteristic speeds increase dramatically (see Fig.~\ref{fig:vcharconfronto}), and the chosen time step is no longer small enough to ensure stability.
Indeed, a simulation performed with $\mathrm{CFL} = 0.0625$ exhibits a rapid growth of the Kretschmann scalar up to $K \sim 10^{11}\tilde{M}^{-4}$, signaling numerical breakdown, as shown in the left panel of Fig.~\ref{fig:K_A5_highCFL_breakNobreak} and also by the purple dashed curve in Fig.~\ref{fig:vcharconfronto}. 

Reducing the CFL factor to $0.015625$ restores a regular evolution, as shown in the right panel of Fig.~\ref{fig:K_A5_highCFL_breakNobreak} (see also the green curve in Fig.~\ref{fig:vcharconfronto}). The wave packet now reaches the origin very early ($t \simeq 0.5\tilde{M}$), followed by a bounce and rarefaction phase that generates multiple outgoing curvature pulses. 

A closer inspection reveals that the repulsive contribution of the negative energy density becomes dominant almost immediately: part of the wave packet is prevented from reaching the center and propagates outward instead. The curvature remains finite ($K \lesssim 10^3\tilde{M}^{-4}$) and gradually decays over time. 
 
In all simulations performed with the \textit{phantom} field, the Misner-Sharp mass remains consistently negative throughout the evolution. For matter satisfying the dominant energy condition, $M_{\rm MS}$ is nondecreasing with the areal radius $r$ in an untrapped region~\cite{Hayward:1994bu}. However, this property does not hold for the phantom field. The negativity of the Misner-Sharp mass should therefore be interpreted with care and is consistent with the repulsive character of the field, which violates the null and strong energy conditions. 

\begin{figure}
    \centering
    \includegraphics[width=\linewidth]{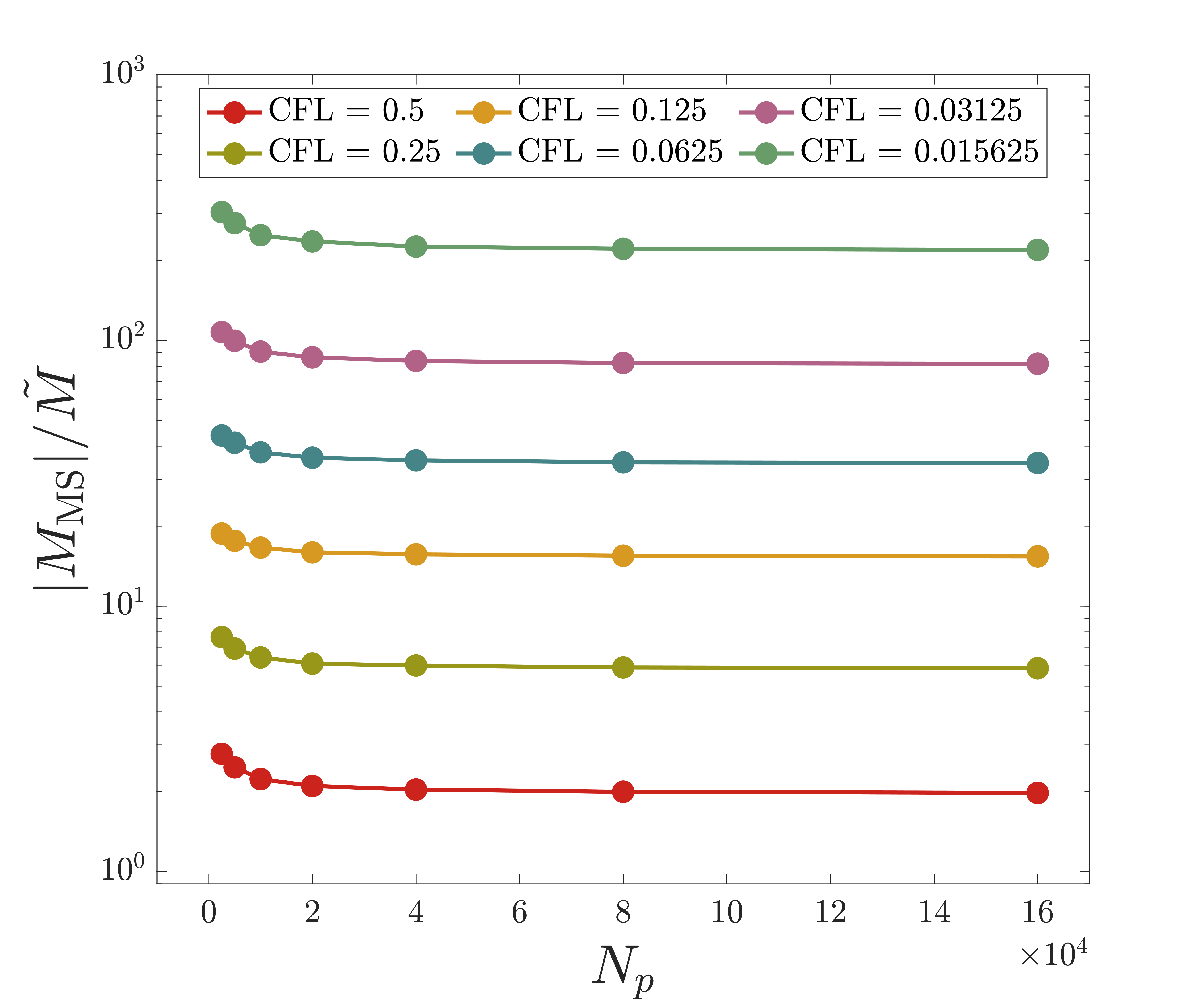}
    \caption{Absolute value of the Misner-Sharp mass, $|M_{\rm MS}|$, computed at the critical amplitude $A_\text{break}$ as a function of the number of radial grid points $N_p$, for different values of the CFL factor.}
    \label{fig:Mms gauss}
\end{figure}

Figure~\ref{fig:Mms gauss} shows the absolute value $|M_{\rm MS}|$ at the critical amplitude $A_\text{break}$ displayed in Fig.~\ref{fig:Astar_N}, plotted as a function of the spatial resolution and the CFL factor.
Even for absolute values of the Misner-Sharp mass as large as $|M_{\rm MS}| \approx 200\tilde{M}$, obtained in runs with smaller CFL factors, the evolution remains regular. For comparison, in the case of a canonical scalar field such a mass would correspond to a black hole with areal radius $r_\text{H} = 2M \approx 400\tilde{M}$. In the present case, however, no trapped region forms and the field ultimately disperses.
Overall, within the explored parameter range, our results show no evidence for the formation of trapped surfaces, naked singularities, or alternative exotic configurations during the collapse of the phantom field.

\subsection{Further Analysis of the Numerical Breakdown}
\label{sec:breakdown}     

Our analysis so far has focused on a specific wave-packet profile. We now generalize it by considering different initial configurations for the phantom field, in order to test the robustness of the observed behavior and assess whether the spatial distribution of the energy affects the outcome.

\begin{figure}
    \centering
    \includegraphics[width=\linewidth]{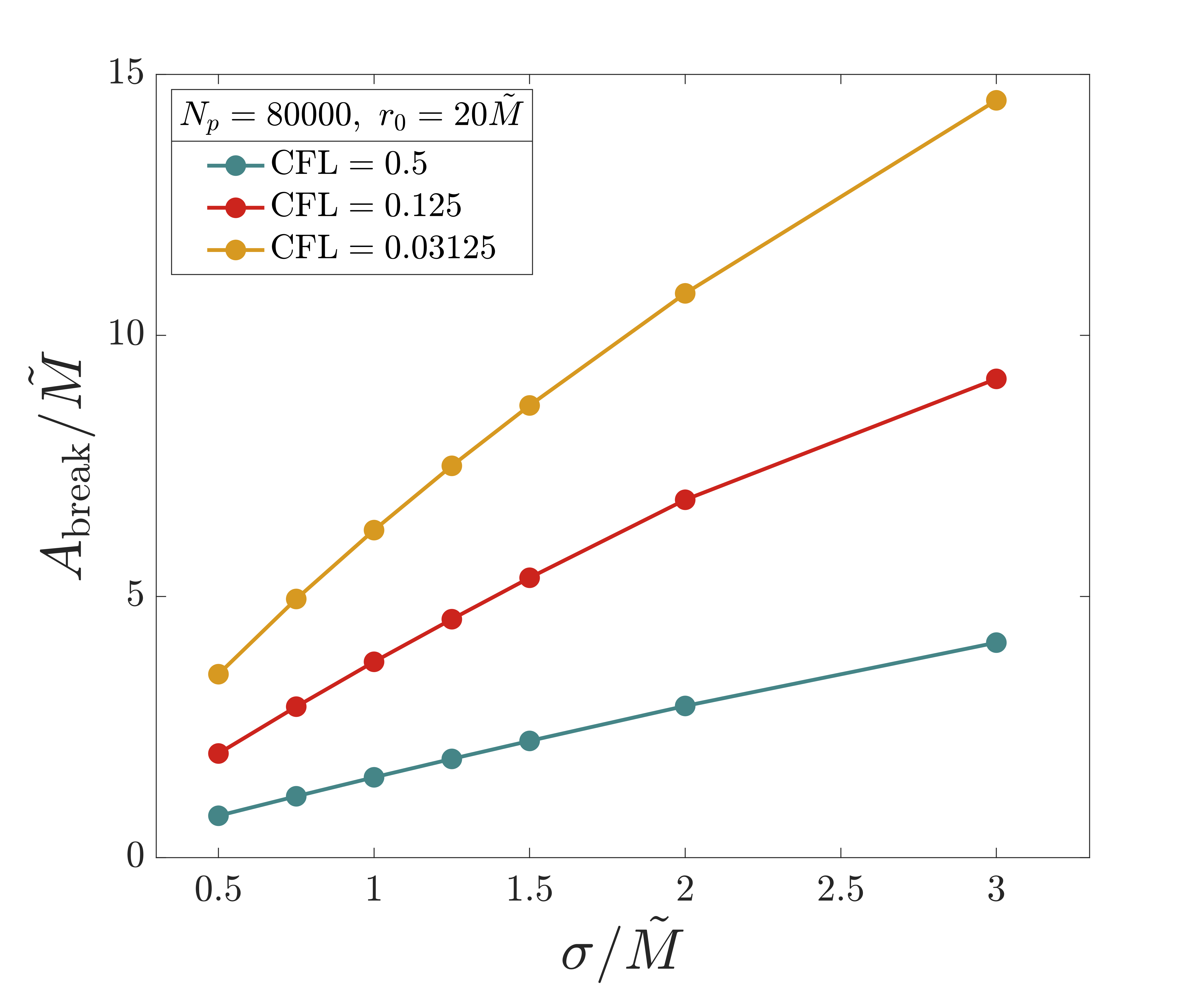}
    \caption{Critical threshold $A_{\rm break}$ as a function of the width $\sigma$, for different values of the CFL. The simulations are performed with $N_p = 80000$ and $r_0 = 20\tilde{M}$.}
    \label{fig:A vs sigma}
\end{figure}

We start by varying the parameters of the Gaussian profile in Eq.~\eqref{gauss pacchetto collasso}: we fix the center of the wave packet at $r_0 = 20\tilde{M}$ and consider different values of the width parameter $\sigma$. 
In this additional set of simulations we adopt $N_p = 80000$, which yields estimates of the critical amplitude close to the asymptotic value at fixed CFL (cf.\ Fig.~\ref{fig:Astar_N}). The results are summarized in Fig.~\ref{fig:A vs sigma}, where we plot the critical amplitude $A_\text{break}$ as a function of $\sigma$ for several choices of the CFL factor.

We observe that $A_\text{break}$ increases monotonically with the width of the wave packet. This trend can be qualitatively understood from the structure of the stress-energy tensor: the effective energy and momentum densities depend on derivatives of $\xi$. A broader wave packet has smaller spatial gradients, which reduce the magnitude of these derivative terms. Consequently, a larger amplitude is required to produce comparable curvature effects and approach the regime where the evolution becomes numerically problematic.

However, the separation between the curves corresponding to different CFL values indicates the absence of convergence. As the time step is reduced, the threshold shifts toward progressively larger amplitudes, with no sign of saturation. This behavior is consistent with our previous analyses and reinforces the conclusion that no physically meaningful, convergent value of $A_\text{break}$ exists, and that the breakdown is driven by numerical rather than dynamical effects.

Lastly, we test the robustness of our results against a qualitatively different initial profile for the phantom field, namely the double-$\tanh$ configuration defined in Eq.~\eqref{tanh xi}. We set ${\hat r}_1 = 10\tilde{M}$, ${\hat r}_2 = 30\tilde{M}$, and $\sigma = \tilde{M}$, choices that ensure consistency with the boundary conditions of the computational domain. From a physical standpoint, the Gaussian and double-$\tanh$ initial configurations represent qualitatively different perturbations.
The Gaussian packet represents a localized and smooth excitation with narrow-band spectral content: its energy is concentrated around a dominant frequency, and the collapse proceeds in a relatively simple and temporally coherent manner. By contrast, the double-$\tanh$ profile describes an extended configuration featuring quasi-stationary regions separated by steep gradients at the edges. These sharp transitions introduce higher-frequency components and a slower spectral decay, resulting in a broader distribution of scales.

In dynamical terms, this distinction translates into a more intricate time evolution. While the Gaussian perturbation typically propagates inward and either disperses or collapses within a finite time, the step-like configuration evolves more gradually, producing successive wavefronts and a less regular temporal pattern. This setup therefore provides a complementary probe of the dynamics, allowing us to assess how sharper gradients and multiple time scales affect the gravitational focusing properties of the phantom field. 

\begin{figure}
    \centering
    \includegraphics[width=\columnwidth]{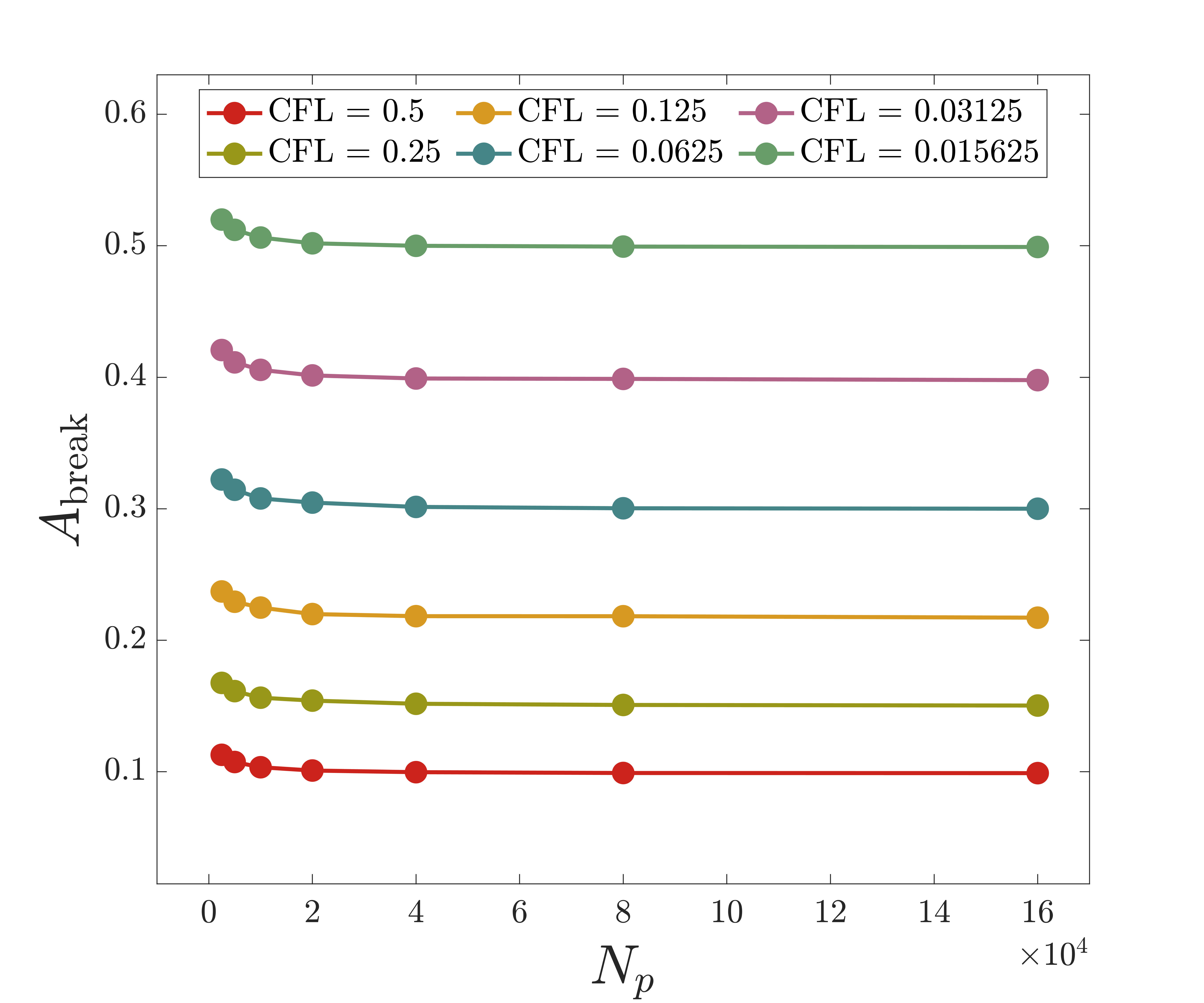} \\
    \includegraphics[width=\columnwidth]{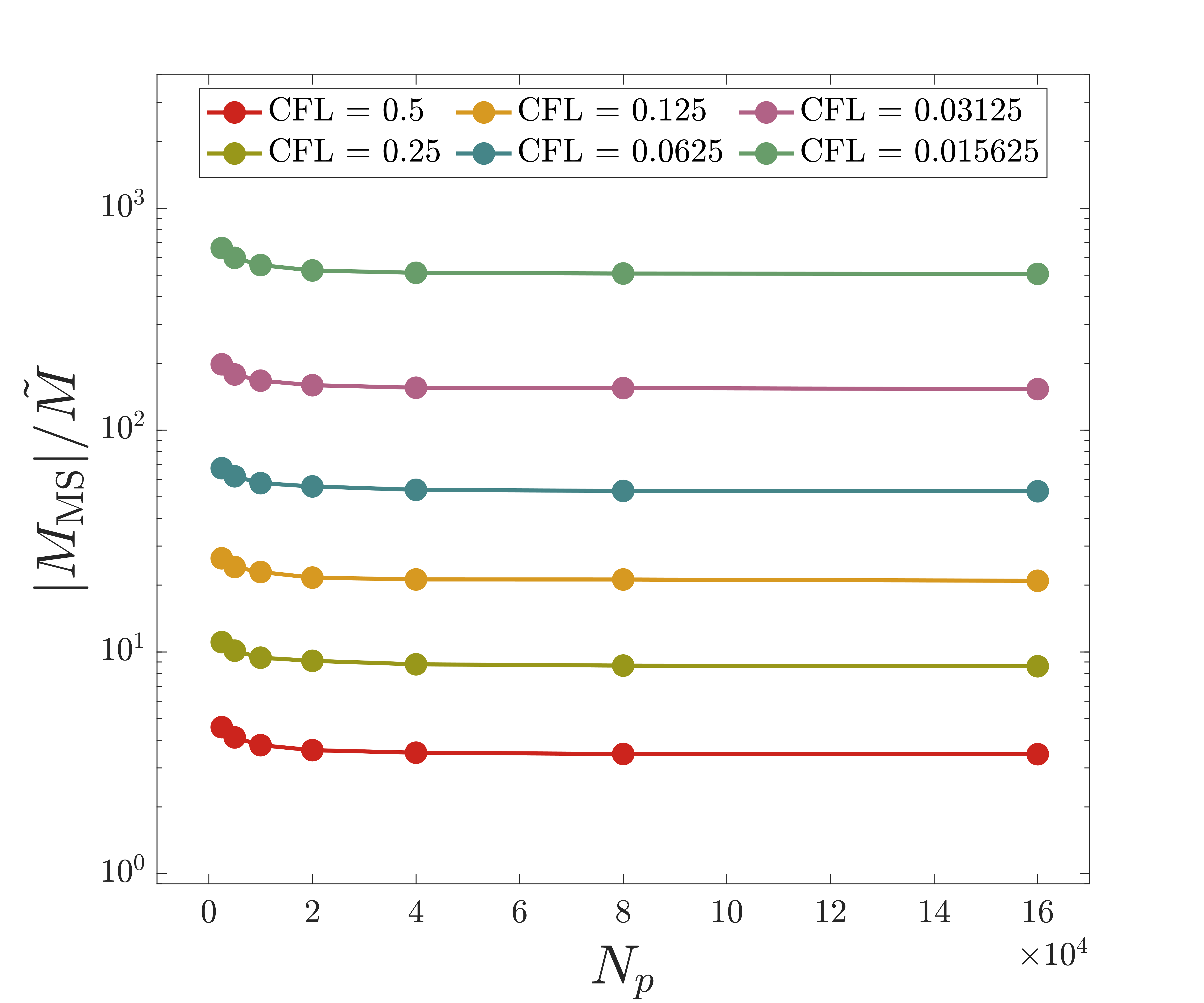}
    \caption{Threshold of numerical breakdown as a function of the number of grid points, for different choices of the CFL factor, in the case of an initial phantom wave packet of tanh-type profile, Eq.~\eqref{tanh xi}. The upper panel shows the values of the critical amplitude $A_\text{break}$, and the lower panel the corresponding Misner-Sharp mass of the system (in absolute value). Both quantities show no clear convergence trend as the CFL decreases, confirming that the numerical breakdown is dominated by the choice of the time step, rather than by spatial resolution.}
    \label{fig:confronto tanh}
\end{figure}

In Fig.~\ref{fig:confronto tanh} we summarize the results obtained with the double-$\tanh$ initial profile. The upper panel shows the breakdown amplitude $A_\text{break}$, while the lower panel displays the absolute value of the Misner-Sharp mass at the onset of the breakdown, for different spatial resolutions and CFL factors.
The qualitative behavior is fully consistent with that found for the Gaussian profile. In particular, we observe clear convergence of the critical threshold $A_\text{break}$ with increasing spatial resolution, whereas no analogous convergence emerges with respect to the CFL factor. 

The present analysis therefore supports the conclusion that the observed breakdown is a numerical artifact and that it is predominantly controlled by the choice of time step rather than by the physical properties of the initial configuration.

\section{Conclusions}  \label{sec:conclusions}
In this work we have investigated the nonlinear gravitational collapse of a phantom scalar field in asymptotically flat spacetime. Because the kinetic term of the phantom field enters the action with the opposite sign relative to a canonical scalar field, its energy density can become negative. This raises the possibility that gravitational collapse could lead to exotic outcomes, such as negative-mass configurations, naked singularities, or other violations of the cosmic censorship conjecture.

To address this \textit{gedanken experiment}, we performed high-accuracy numerical-relativity simulations of spherically symmetric configurations, evolving smooth and regular wave packets of the phantom scalar field across a wide range of amplitudes. The numerical scheme was validated through convergence tests and monitoring of the constraint equations.

Our results show that none of the initial data considered lead to the formation of trapped surfaces or singular end states. Instead, the scalar field disperses and the spacetime relaxes toward a regular asymptotically flat geometry. In particular, we find no evidence for the formation of negative-mass Schwarzschild solutions or other stationary exotic remnants. Within the parameter space explored here, the nonlinear dynamics therefore appears to prevent the realization of configurations that would violate cosmic censorship.

All simulations were performed by solving the Einstein-Klein-Gordon system in spherical symmetry, discretized using the method of lines and evolved with a fourth-order Runge-Kutta integrator. We considered two classes of initial data, a Gaussian pulse and a double-$\tanh$ profile, adopting a Choptuik-like strategy in which the shape and width were fixed while the amplitude was varied. In both families, increasing the temporal resolution removes the apparent numerical breakdown: rather than collapsing, the scalar field ultimately disperses.

The breakdown observed at large amplitudes is therefore purely numerical. It arises when strong metric gradients cause the local characteristic speeds to grow significantly. Under these conditions the Courant-Friedrichs-Lewy stability criterion is violated, triggering numerical instabilities. Reducing the CFL factor restores stable and regular evolution.
We further analyzed the nonlinear dynamics through the evolution of the Kretschmann scalar, which provides a clear diagnostic of curvature amplification, wave focusing, and subsequent dispersion. Even for configurations corresponding to effective masses as negative as $M_{\rm MS} \sim -200 \tilde M$, the curvature remains finite and no trapped surfaces form. For comparison, a canonical scalar field with similar parameters would produce black holes with macroscopic horizons.

Within the parameter space explored here, we find no indication of singularity formation, horizon development, or alternative stationary solutions such as wormhole-like geometries. Instead, the repulsive character of the phantom field systematically counteracts gravitational focusing and drives the system toward a regular dispersive state, consistently with the behavior reported in Refs.~\cite{Nakonieczna:2012in,Nakonieczna:2013hs}.

The difficulties we encountered from the numerical point of view are mainly due to the choice of Schwarzschild-like coordinates, which produce huge jumps in the characteristic velocities when considering large amplitudes, and forced us to significantly decrease the time step. While a different gauge choice could help in making the simulation more stable, and allow to simulate the same scenario at a lower computational cost, the qualitative outcome of our work, i.e. the dispersion of the field for all the initial configurations considered, is expected to be independent of the coordinate choice. However, an independent numerical evolution using horizon-penetrating coordinates would provide further confirmation of our results and is left for future work.

The numerical framework developed in this work proves robust and well controlled, successfully handling the highly dynamical regime associated with large characteristic velocities. Although the simulations require fine spatial and temporal resolution, their convergence properties and stability behavior are well understood.

Overall, our results indicate that cosmic censorship appears to remain dynamically preserved even in the presence of negative-energy matter.

\begin{acknowledgements}
    This research was supported by the MUR FIS2 Advanced Grant ET-NOW (CUP:~B53C25001080001) and by the INFN TEONGRAV initiative. 
    Numerical computations have been performed at the Vera and CHRONOS clusters supported by the Italian Ministry of Research and by Sapienza University of Rome.
\end{acknowledgements}

\appendix

\section{Code validation}
\label{app:convergence}

To quantify the accuracy of the nonlinear $1+1$ code, we check how the violation of the radial constraint for $\beta$ (Eq.~\eqref{eq:dr_B}) scales with resolution. Such constraint violation (CV) is evaluated as
\begin{equation}
    CV_\beta(t,r)
    =
    \partial_r \beta
    +
    \frac{e^{2 \beta}\!\left(C r^2 \Pi^2 + 1\right) + C r^2 \Theta^2 - 1}{2r},
\end{equation}
where $\partial_r \beta$ is computed numerically.

Since both spatial discretization and time integration are implemented with fourth-order schemes, halving the spatial grid spacing is expected to reduce the constraint violation according to the relation
\begin{align}
CV_{\Delta r_{\text{coarse}}}
=
\left(
\frac{\Delta r_{\text{coarse}}}{\Delta r_{\text{fine}}}
\right)^4
CV_{\Delta r_{\text{fine}}}
=
16\, CV_{\Delta r_{\text{fine}}},
\label{cv scaling low}
\end{align}
where $\Delta r_{\text{coarse}} = 2 \Delta r_{\text{fine}}$.  
Therefore, if the implementation is correct, the fine-grid CV multiplied by 16 should overlap with the coarse-grid CV.

We perform convergence tests using pairs of simulations with spatial resolutions differing by a factor two (e.g., $\Delta r_{\text{coarse}}=0.01\tilde{M}$, $\Delta r_{\text{fine}}=0.005\tilde{M}$). Each pair is evolved with various CFL values to separately assess the effects of spatial and temporal resolution.

\begin{figure}
\centering
\includegraphics[width=\linewidth]{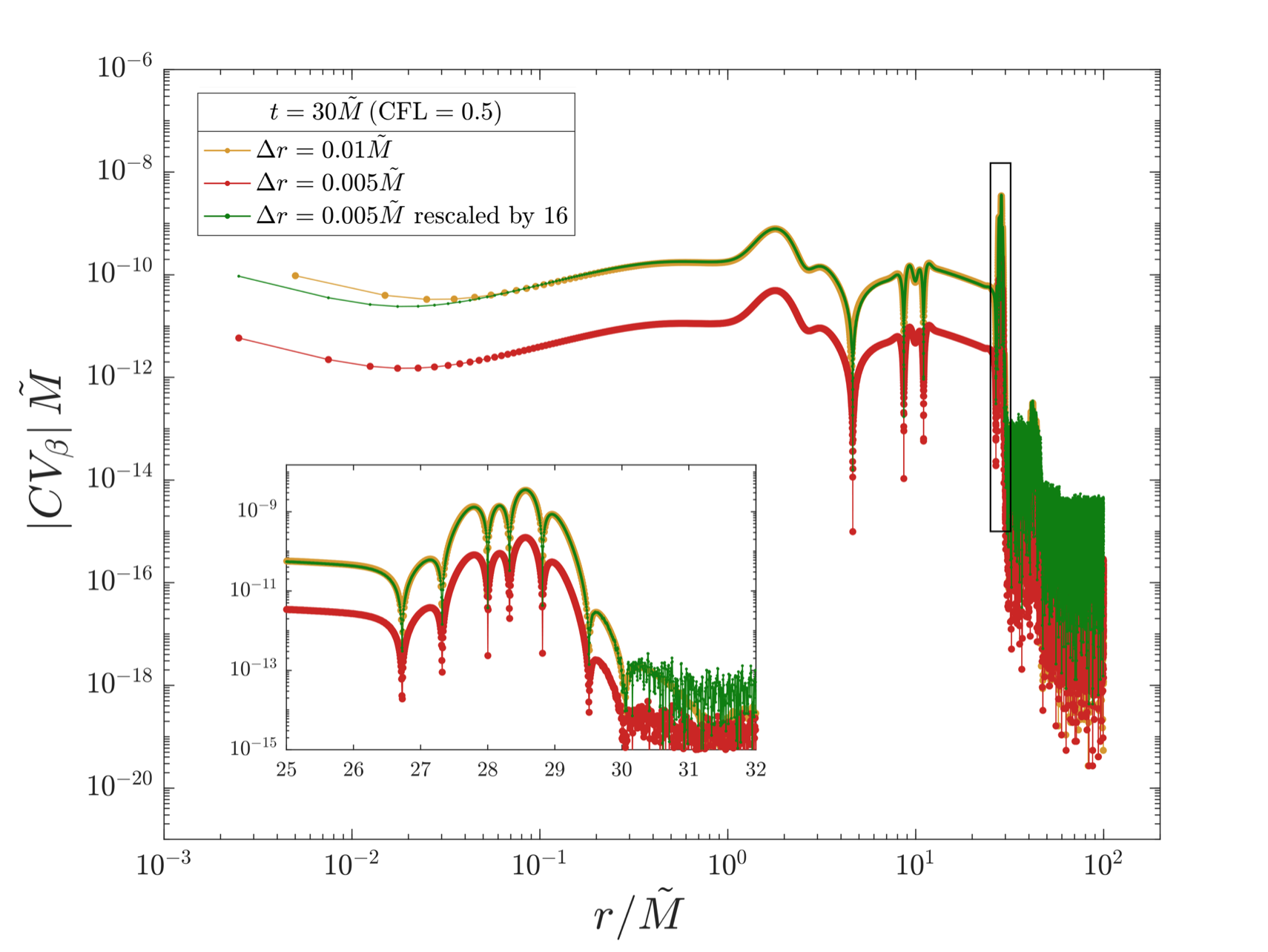}
\caption{Fourth-order scaling of the constraint violation $CV_\beta$ at $t=30\tilde{M}$, with initial data $(A,\sigma,r_0)=(1.0,1.0,10)\tilde{M}$ and $\mathrm{CFL}=0.5$. 
Yellow: $\Delta r=0.01\tilde{M}$; red: $\Delta r=0.005\tilde{M}$; green: fine grid data rescaled by 16.}
\label{fig:CV_collasso1}
\end{figure}

First we consider the case of an initial Gaussian wave packet with $A = \tilde M$, $\sigma = \tilde M$ and $r_0 = 10 \tilde M$.
Figure~\ref{fig:CV_collasso1} shows the scaling of $CV_\beta$ at $t=30\tilde{M}$ for $\mathrm{CFL}=0.5$. After rescaling the fine-grid curve by 16, we observe excellent overlap across the entire domain, demonstrating global fourth-order convergence. The inset highlights the region $25 < r/\tilde{M} < 32$, where the scalar field is localized and the violations are largest, yet still obey the expected scaling.

\begin{figure}
\centering
\includegraphics[width=\linewidth]{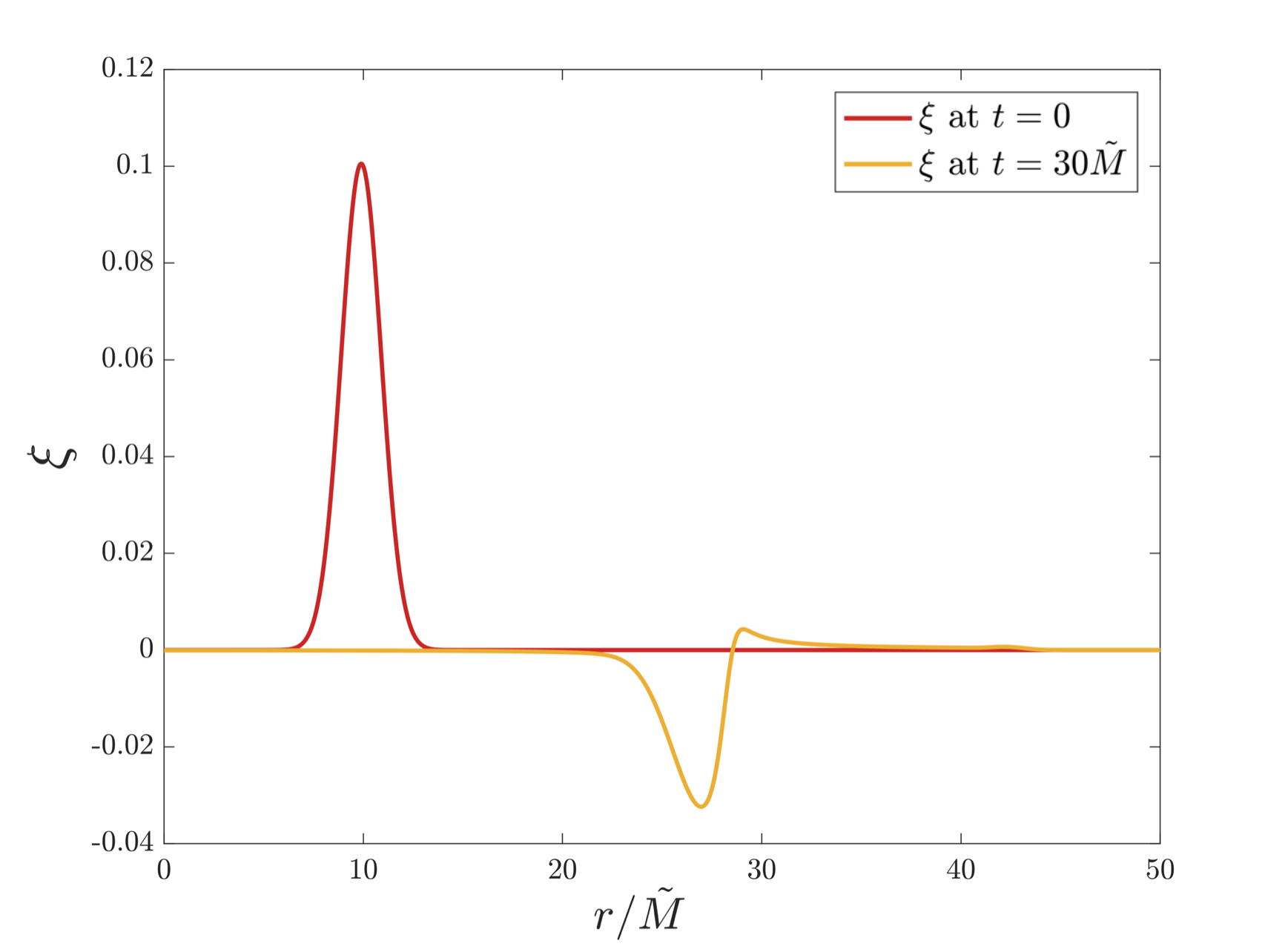}
\caption{Radial profile of $\xi$ for $\Delta r=0.05\tilde{M}$, $(A,\sigma,r_0)=(1.0,1.0,10)\tilde{M}$, and $\mathrm{CFL}=0.5$. 
Red: initial data; yellow: profile at $t=30\tilde{M}$.}
\label{fig:CV_onda}
\end{figure}

To correlate constraint violation with field dynamics, Fig.~\ref{fig:CV_onda} shows the radial profile of $\xi$ at $t=0$ and $t=30\tilde{M}$. The largest violations occur precisely where the field amplitude is significant, as expected since the constraint contains matter terms.

\begin{figure}
\centering
\includegraphics[width=\linewidth]{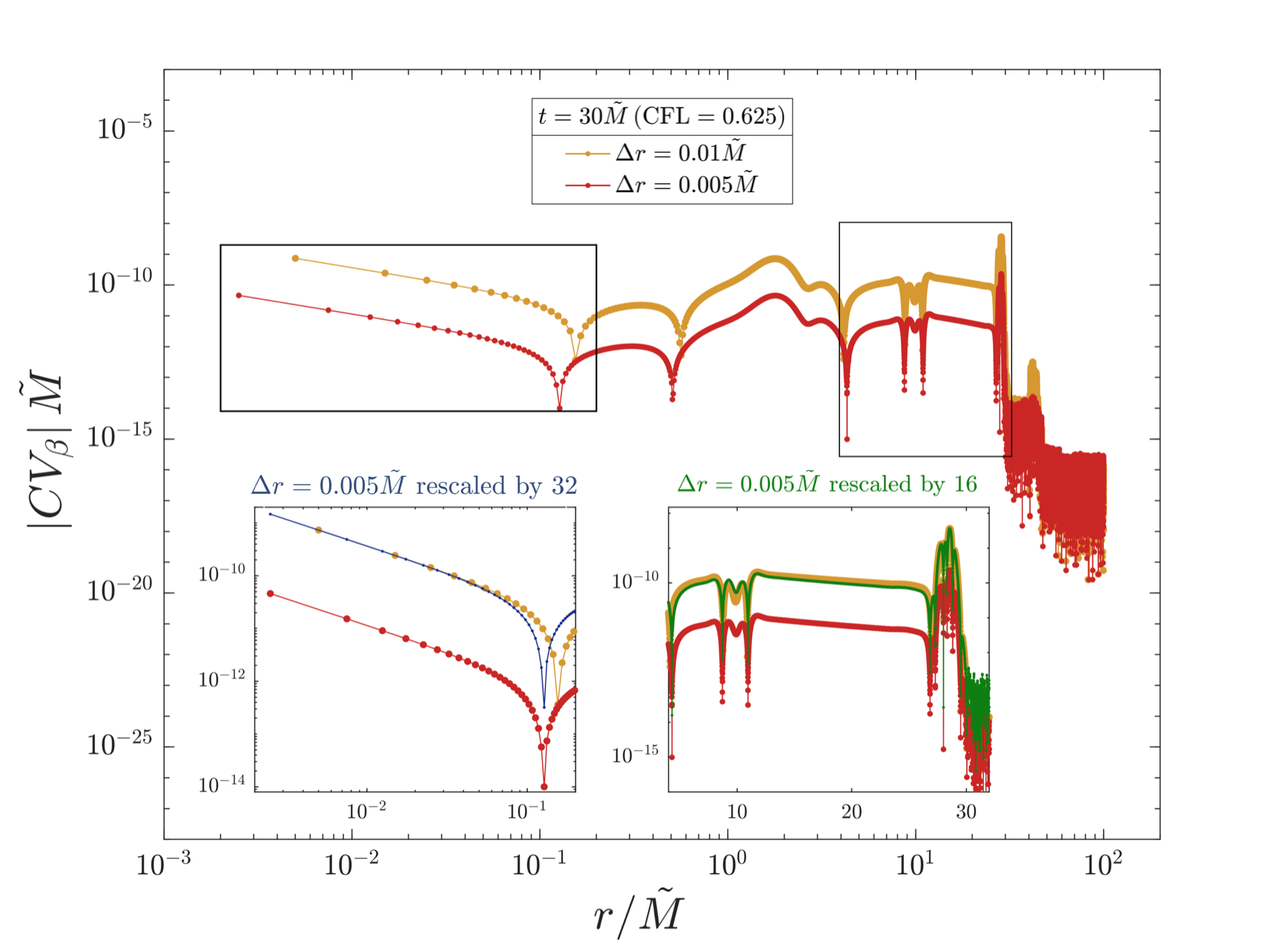}
\caption{Constraint violation $CV_\beta$ at $t=30\tilde{M}$ for $\mathrm{CFL}=0.0625$. 
Yellow: $\Delta r=0.01\tilde{M}$; red: $\Delta r=0.005\tilde{M}$.}
\label{fig:CV_collasso2}
\end{figure}

Next, we repeat the analysis for smaller CFL values. Figure~\ref{fig:CV_collasso2} shows $CV_\beta$ at $t=30\tilde{M}$ for $\mathrm{CFL}=0.0625$. Fourth-order convergence is again observed across most of the domain when rescaling by 16. In localized regions, improved overlap is obtained with a factor 32. This behavior is attributable to the fifth-order Kreiss-Oliger dissipation term~\eqref{eq:KO}, which modifies the highest-frequency components and slightly alters the effective convergence order in smooth regions.

\begin{figure}
\centering
\includegraphics[width=\linewidth]{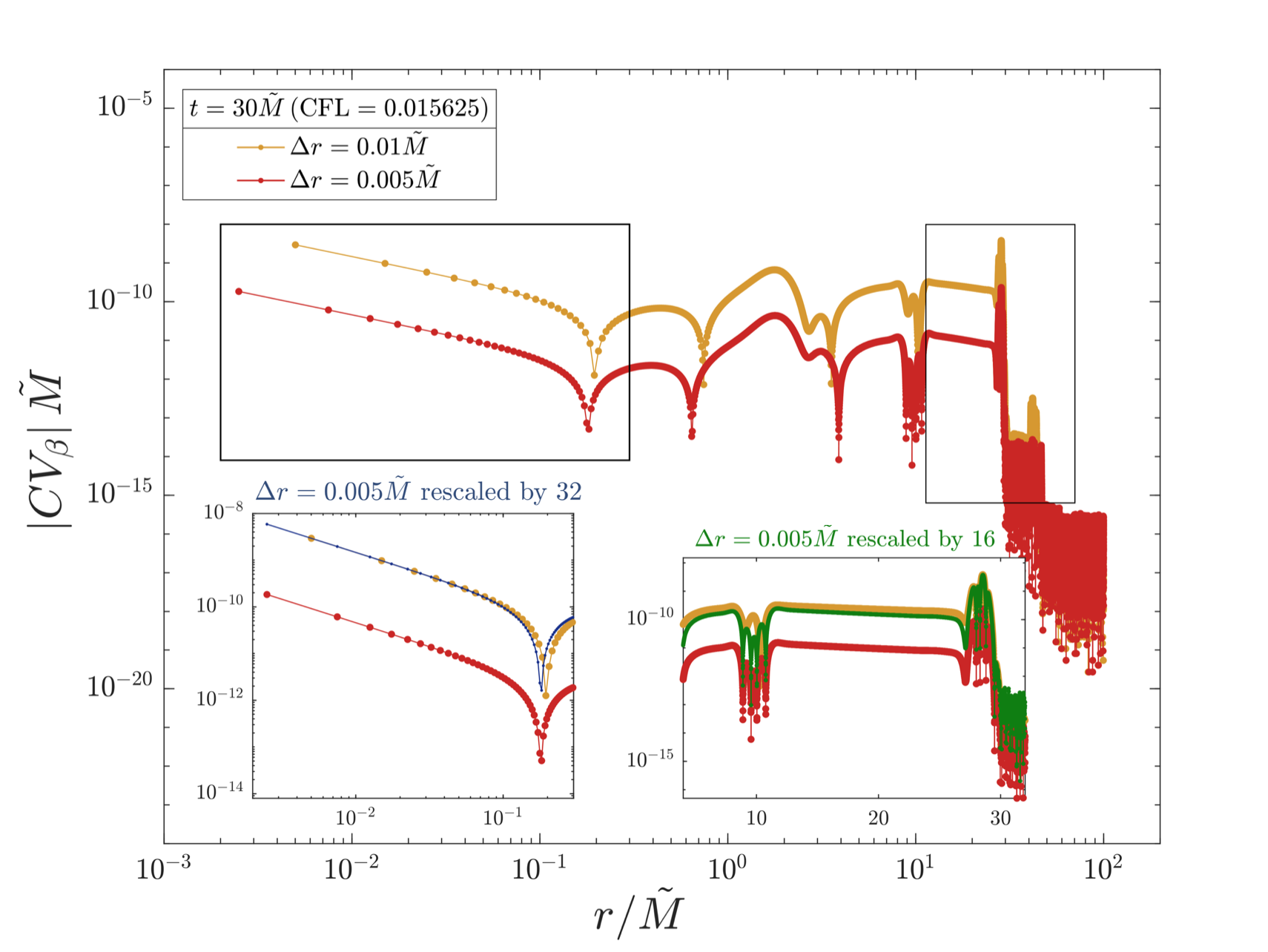}
\caption{Constraint violation $CV_\beta$ at $t=30\tilde{M}$ for $\mathrm{CFL}=0.015625$.}
\label{fig:CV_collasso3}
\end{figure}

A similar behavior is observed for $\mathrm{CFL}=0.015625$, as shown in Fig.~\ref{fig:CV_collasso3}. At late times ($t=30\tilde{M}$), mild deviations from ideal fourth-order scaling appear more prominently. Such deviations are consistent with an increased relative influence of the dissipation term once the field amplitude has decreased and the solution becomes smoother.

\begin{figure}
\centering
\includegraphics[width=\linewidth]{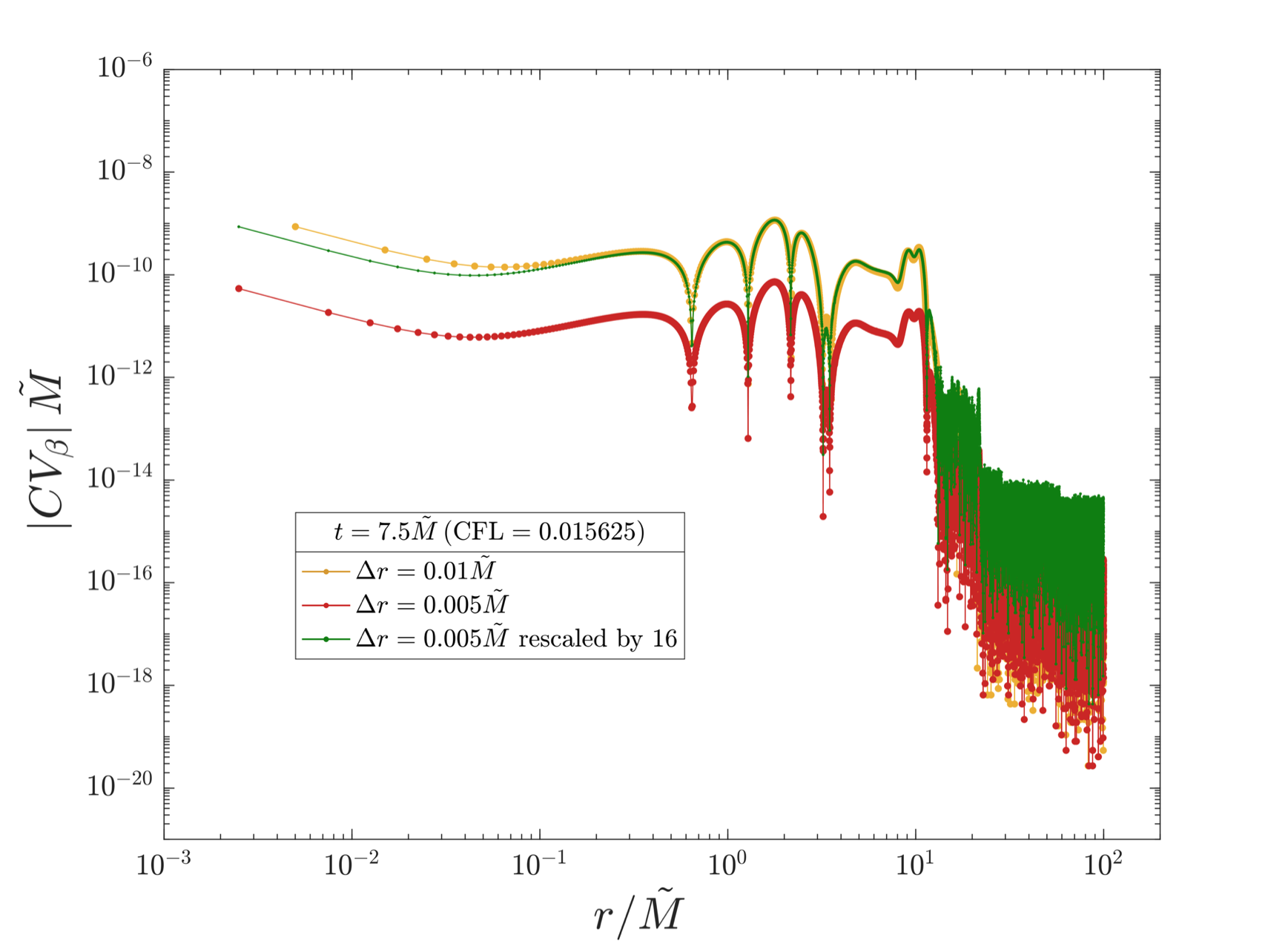}
\caption{Constraint violation $CV_\beta$ at $t=7.5\tilde{M}$ for $\mathrm{CFL}=0.015625$.}
\label{fig:CV_collasso4}
\end{figure}

To corroborate this hypothesis, we examined earlier times when nonlinear dynamics are stronger and the dissipative term is subdominant. At $t=7.5\tilde{M}$, shortly before the collapse phase, fourth-order scaling is clearly recovered even for $\mathrm{CFL}=0.015625$, as shown in Fig.~\ref{fig:CV_collasso4}.

Furthermore, to verify the robustness of the boundary treatment (see Sec.~\ref{sec:NumericalIntegration}), which employs a half-step staggered grid together with three ghost cells to enforce the parity conditions, we examine the constraint violation $CV_\beta$ at the time of the bounce for a high-amplitude pulse ($A=5\tilde{M}$). In this regime, the spatial gradients of $\Theta=\partial_r\xi$ become extremely steep and could potentially compromise the fourth-order accuracy of the numerical scheme.

\begin{figure}
    \centering
    \includegraphics[width=\linewidth]{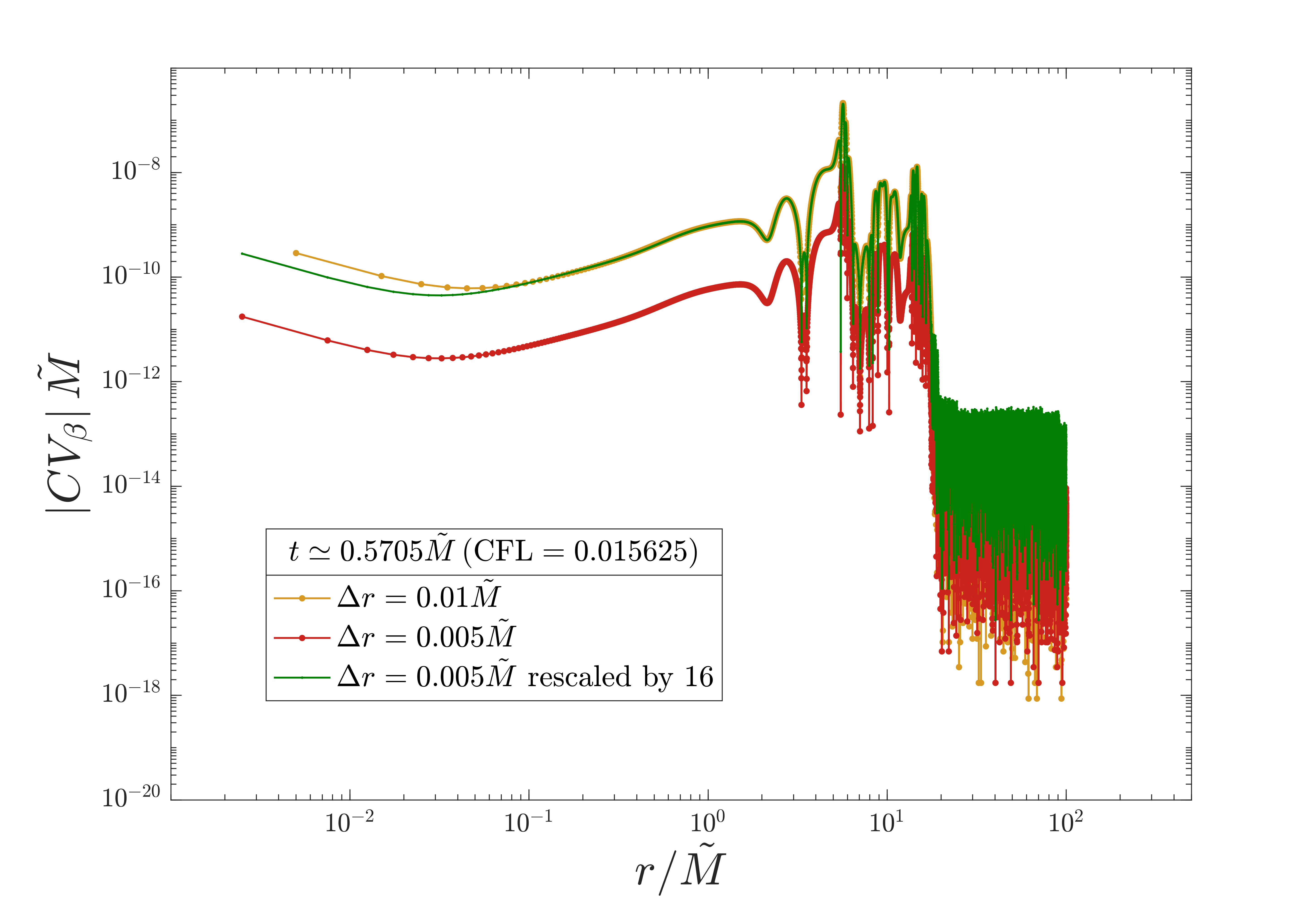}
    \caption{Constraint violation $CV_\beta$ at $t\simeq 0.5705\tilde{M}$, corresponding to the moment of maximum compression (bounce) for a simulation with $A=5\tilde{M}$ and $\mathrm{CFL}=0.015625$. The fine-grid curve (green), rescaled by a factor of 16, demonstrates fourth-order convergence at the bounce.}
    \label{fig:CVbounce}
\end{figure}

However, as shown in Fig.~\ref{fig:CVbounce}, no loss of accuracy is observed: the constraint remains well behaved, and fourth-order convergence is preserved.
Overall, these tests demonstrate that the numerical implementation achieves the expected fourth-order convergence and that deviations at late times are consistent with the influence of Kreiss-Oliger dissipation rather than with deficiencies in the underlying scheme.

\section{Impact of the dissipation operator on the numerical breakdown}
\label{app:KreissOligerImpact}

To assess the influence of the Kreiss-Oliger dissipation operator on our results, we carried out an additional analysis in which the dissipation parameter $\eta_{\rm KO}$ was varied while keeping $r_0=10\tilde{M}$ and $\sigma=\tilde{M}$ fixed. The results are collected in Table~\ref{tab:Astar_cfl}, whose different columns give the values of $A_\text{break}$ for several choices of the CFL factor and spatial resolution. We find that $A_\text{break}$ shows only a weak dependence on $\eta_{\rm KO}$, with variations of at most a few $\times 10^{-2}$, while its dependence on the numerical resolution and, especially, on the time step is substantially stronger.
These results indicate that the determination of the critical threshold is insensitive to reasonable variations of the numerical dissipation and further support the conclusion that the observed breakdown is primarily associated with the temporal discretization.

\begin{table}
\renewcommand{\arraystretch}{1.2}
    \centering
   \caption{Critical amplitude $A_\text{break}$ as a function of the dissipation parameter $\eta_{\rm KO}$, computed at different spatial resolutions and CFL factors. Variations in $A_\text{break}$ with $\eta_{\rm KO}$ are at the percent level, whereas the dependence on the CFL factor is significantly stronger.}
    \begin{ruledtabular}
        \begin{tabular}{dddd}
            \multicolumn{1}{c}{\multirow{2}{*}{\diagbox{$\eta_{\rm KO}$}{$\frac{A_\text{break}}{\tilde{M}}$}}} & \multicolumn{1}{c}{$\text{CFL}=0.5$}  & \multicolumn{1}{c}{$\rm CFL=0.0625$} & \multicolumn{1}{c}{$\rm CFL=0.015625$} \\
            & \multicolumn{1}{c}{$N_p=40000$} & \multicolumn{1}{c}{$N_p=80000$} & \multicolumn{1}{c}{$N_p=160000$}\\
            \colrule
            0.025 & 1.442 & 4.138 & 5.989 \\
            0.05  & 1.446 & 4.139 & 5.992 \\
            0.1   & 1.448 & 4.143 & 5.996 \\
            0.2   & 1.453 & 4.150 & 6.003 \\
            0.4   & 1.466 & 4.163 & 6.019 \\
        \end{tabular}
    \end{ruledtabular}
    \label{tab:Astar_cfl}
\end{table}


\bibliographystyle{apsrev4-1}
\bibliography{ref.bib}

\end{document}